\pdfoutput=1
\documentclass[12pt,a4paper]{article}

\usepackage{ifthen} 
\newboolean{pdflatex}
\setboolean{pdflatex}{true} 

\newboolean{articletitles}
\setboolean{articletitles}{true} 
\usepackage{adjustbox}
\newboolean{uprightparticles}
\setboolean{uprightparticles}{false} 

\def\paperauthors{LHCb collaboration} 

\def\paperasciititle{Improved branching-fraction measurements of $B^0_{(s)} \to K_S^0 h^+ h^{'-}$ decays and first observation of $B^0_s \to K_S^0 K^+ K^-$} 
\def\papertitle{Improved branching-fraction measurements of $B^0_{(s)} \to K_{\mathrm{\scriptscriptstyle S}}^0 h^+ h^{\prime -}$ decays and first observation of $B_s^0 \to K_{\mathrm{\scriptscriptstyle S}}^0 K^+ K^-$} 

\def\paperkeywords{{High Energy Physics}, {LHCb}} 
\def\papercopyright{\the\year\ CERN for the benefit of the LHCb collaboration} 
\def\paperlicence{CC BY 4.0 licence}
\def\paperlicenceurl{https://creativecommons.org/licenses/by/4.0/}

\newif\ifEnableSectionTOCLinks
\EnableSectionTOCLinksfalse 

\usepackage[top=1in, bottom=1.25in, left=1in, right=1in]{geometry}

\usepackage{microtype}
\usepackage{lineno}  
\usepackage{xspace} 
\usepackage{caption} 

\usepackage{graphicx}  
\usepackage{color}
\usepackage{colortbl}
\graphicspath{{./figs/}} 

\usepackage{amsmath} 
\usepackage{amssymb}
\usepackage{amsfonts}
\usepackage{upgreek} 

\newcommand*\patchAmsMathEnvironmentForLineno[1]{%
\expandafter\let\csname old#1\expandafter\endcsname\csname #1\endcsname
\expandafter\let\csname oldend#1\expandafter\endcsname\csname
end#1\endcsname
 \renewenvironment{#1}%
   {\linenomath\csname old#1\endcsname}%
   {\csname oldend#1\endcsname\endlinenomath}%
}
\newcommand*\patchBothAmsMathEnvironmentsForLineno[1]{%
  \patchAmsMathEnvironmentForLineno{#1}%
  \patchAmsMathEnvironmentForLineno{#1*}%
}
\AtBeginDocument{%
\patchBothAmsMathEnvironmentsForLineno{equation}%
\patchBothAmsMathEnvironmentsForLineno{align}%
\patchBothAmsMathEnvironmentsForLineno{flalign}%
\patchBothAmsMathEnvironmentsForLineno{alignat}%
\patchBothAmsMathEnvironmentsForLineno{gather}%
\patchBothAmsMathEnvironmentsForLineno{multline}%
\patchBothAmsMathEnvironmentsForLineno{eqnarray}%
}

\usepackage[pdftex,
            pdfauthor={\paperauthors},
            pdftitle={\paperasciititle},
            pdfkeywords={\paperkeywords}]{hyperref}

\usepackage{hyperxmp}

\hypersetup{
            pdfcopyright={Copyright (C) \papercopyright},
            pdflicenseurl={\paperlicenceurl}
}
\usepackage[colorinlistoftodos,textsize=scriptsize]{todonotes}

\usepackage[bottom,flushmargin,hang,multiple]{footmisc}

\usepackage[all]{hypcap} 

\usepackage{booktabs}

\usepackage{xspace} 
\usepackage{upgreek}

\def\lhcb   {\mbox{LHCb}\xspace}

\def\babar  {\mbox{BaBar}\xspace}

\def\velo   {VELO\xspace}

\def\MagUp {\mbox{\em Mag\kern -0.05em Up}\xspace}

\ifthenelse{\boolean{uprightparticles}}%
{
 
 \def\Pgamma      {\ensuremath{\upgamma}\xspace}

 \def\Peta        {\ensuremath{\upeta}\xspace}

 \def\Ppi         {\ensuremath{\uppi}\xspace}

 \def\Pchi        {\ensuremath{\upchi}\xspace}                 
 \def\Ppsi        {\ensuremath{\uppsi}\xspace}

 \def\PDelta      {\ensuremath{\Delta}\xspace}                 
 \def\PXi         {\ensuremath{\Xi}\xspace}                 
 \def\PLambda     {\ensuremath{\Lambda}\xspace}                 
 \def\PSigma      {\ensuremath{\Sigma}\xspace}                 
 \def\POmega      {\ensuremath{\Omega}\xspace}                 
 \def\PUpsilon    {\ensuremath{\Upsilon}\xspace}
 \let\oldPi\Pi
 \def\PPi         {\ensuremath{\oldPi}\xspace}

 \def\PB      {\ensuremath{\mathrm{B}}\xspace}                 
 \def\PD      {\ensuremath{\mathrm{D}}\xspace}                 
 \def\PJ      {\ensuremath{\mathrm{J}}\xspace}                 
 \def\PK      {\ensuremath{\mathrm{K}}\xspace}                 
 \def\Pb      {\ensuremath{\mathrm{b}}\xspace}                 
 \def\Pc      {\ensuremath{\mathrm{c}}\xspace}                 
 \def\Pd      {\ensuremath{\mathrm{d}}\xspace}

 \def\Ph      {\ensuremath{\mathrm{h}}\xspace}                 
 \def\Pp      {\ensuremath{\mathrm{p}}\xspace}                 
 \def\Pq      {\ensuremath{\mathrm{q}}\xspace}                 
 \def\Ps      {\ensuremath{\mathrm{s}}\xspace}

 \def\thebaroffset{0.0em}
}
{
 
 \def\Pgamma      {\ensuremath{\gamma}\xspace}

 \def\Peta        {\ensuremath{\eta}\xspace}

 \def\Ppi         {\ensuremath{\pi}\xspace}

 \def\Pchi        {\ensuremath{\chi}\xspace}                 
 \def\Ppsi        {\ensuremath{\psi}\xspace}                 
                  
 \mathchardef\PDelta="7101
 \mathchardef\PXi="7104
 \mathchardef\PLambda="7103
 \mathchardef\PSigma="7106
 \mathchardef\POmega="710A
 \mathchardef\PUpsilon="7107
 \mathchardef\PPi="7105
 \def\PB      {\ensuremath{B}\xspace}                 
 \def\PD      {\ensuremath{D}\xspace}                 
 \def\PJ      {\ensuremath{J}\xspace}                 
 \def\PK      {\ensuremath{K}\xspace}                 
 \def\Pb      {\ensuremath{b}\xspace}                 
 \def\Pc      {\ensuremath{c}\xspace}                 
 \def\Pd      {\ensuremath{d}\xspace}

 \def\Ph      {\ensuremath{h}\xspace}                 
 \def\Pp      {\ensuremath{p}\xspace}                 
 \def\Pq      {\ensuremath{q}\xspace}                 
 \def\Ps      {\ensuremath{s}\xspace}

 \def\thebaroffset{0.18em}
}
\newcommand{\offsetoverline}[2][\thebaroffset]{\kern #1\overline{\kern -#1 #2}}%

\makeatletter
\ifcase \@ptsize \relax
  \newcommand{\miniscule}{\@setfontsize\miniscule{4}{5}}
\or
  \newcommand{\miniscule}{\@setfontsize\miniscule{5}{6}}
\or
  \newcommand{\miniscule}{\@setfontsize\miniscule{5}{6}}
\fi
\makeatother

\DeclareRobustCommand{\optbar}[1]{\shortstack{{\miniscule (\rule[.5ex]{1.25em}{.18mm})}
  \\ [-.7ex] $#1$}}

\def\quark     {{\ensuremath{\Pq}}\xspace}
\def\quarkbar  {{\ensuremath{\overline \quark}}\xspace}

\def\dquark    {{\ensuremath{\Pd}}\xspace}

\def\squark    {{\ensuremath{\Ps}}\xspace}

\def\cquark    {{\ensuremath{\Pc}}\xspace}
\def\cquarkbar {{\ensuremath{\overline \cquark}}\xspace}

\def\bquark    {{\ensuremath{\Pb}}\xspace}

\def\pion   {{\ensuremath{\Ppi}}\xspace}
\def\piz    {{\ensuremath{\pion^0}}\xspace}
\def\pip    {{\ensuremath{\pion^+}}\xspace}
\def\pim    {{\ensuremath{\pion^-}}\xspace}

\def\pimp   {{\ensuremath{\pion^\mp}}\xspace}

\def\kaon    {{\ensuremath{\PK}}\xspace}
\def\Kbar    {{\ensuremath{\offsetoverline{\PK}}}\xspace}

\def\KorKbar {\kern \thebaroffset\optbar{\kern -\thebaroffset \PK}{}\xspace}
\def\Kz      {{\ensuremath{\kaon^0}}\xspace}
\def\Kzb     {{\ensuremath{\Kbar{}^0}}\xspace}
\def\Kp      {{\ensuremath{\kaon^+}}\xspace}
\def\Km      {{\ensuremath{\kaon^-}}\xspace}
\def\Kpm     {{\ensuremath{\kaon^\pm}}\xspace}

\def\KS      {{\ensuremath{\kaon^0_{\mathrm{S}}}}\xspace}

\def\KorKbarz {\ensuremath{\KorKbar^0}\xspace}

\newcommand{\etapr}{\ensuremath{\Peta^{\prime}}\xspace}

\def\D       {{\ensuremath{\PD}}\xspace}

\def\DorDbar {\kern \thebaroffset\optbar{\kern -\thebaroffset \PD}\xspace}
\def\Dz      {{\ensuremath{\D^0}}\xspace}

\def\Dp      {{\ensuremath{\D^+}}\xspace}
\def\Dm      {{\ensuremath{\D^-}}\xspace}

\def\DpDm    {\ensuremath{\Dp {\kern -0.16em \Dm}}\xspace}

\def\Dstarp  {{\ensuremath{\D^{*+}}}\xspace}

\def\Dsp     {{\ensuremath{\D^+_\squark}}\xspace}

\def\B       {{\ensuremath{\PB}}\xspace}

\def\BorBbar {\kern \thebaroffset\optbar{\kern -\thebaroffset \PB}\xspace}
\def\Bz      {{\ensuremath{\B^0}}\xspace}

\def\Bd      {{\ensuremath{\B^0}}\xspace}

\def\BdorBdbar {\kern \thebaroffset\optbar{\kern -\thebaroffset \Bd}\xspace}

\def\Bs      {{\ensuremath{\B^0_\squark}}\xspace}

\def\BsorBsbar {\kern \thebaroffset\optbar{\kern -\thebaroffset \Bs}\xspace}

\def\jpsi     {{\ensuremath{{\PJ\mskip -3mu/\mskip -2mu\Ppsi}}}\xspace}

\def\chiczero {{\ensuremath{\Pchi_{\cquark 0}}}\xspace}

\def\Y#1S{\ensuremath{\PUpsilon{(#1S)}}\xspace}

\def\proton      {{\ensuremath{\Pp}}\xspace}
\def\antiproton  {{\ensuremath{\overline \proton}}\xspace}

\def\Lz          {{\ensuremath{\PLambda}}\xspace}

\def\LorLbar     {\kern \thebaroffset\optbar{\kern -\thebaroffset \PLambda}\xspace}

\def\Lc          {{\ensuremath{\Lz^+_\cquark}}\xspace}

\def\Lb           {{\ensuremath{\Lz^0_\bquark}}\xspace}

\def\BF         {{\ensuremath{\mathcal{B}}}\xspace}

\newcommand{\decay}[2]{\mbox{\ensuremath{#1\!\to #2}}\xspace} 

\def\to                 {\ensuremath{\rightarrow}\xspace}

\def\order   {{\ensuremath{\mathcal{O}}}\xspace}

\def\eps   {{\ensuremath{\varepsilon}}\xspace}

\def\CP                {{\ensuremath{C\!P}}\xspace}

\newcommand{\DGs}{{\ensuremath{\Delta\Gamma_{\squark}}}\xspace}

\newcommand{\Gs}{{\ensuremath{\Gamma_{\squark}}}\xspace}

\def\AT#1     {\ensuremath{A_{\mathrm{T}}^{#1}}\xspace}           

\def\C#1      {\ensuremath{\mathcal{C}_{#1}}\xspace}                       
\def\Cp#1     {\ensuremath{\mathcal{C}_{#1}^{'}}\xspace}                    
\def\Ceff#1   {\ensuremath{\mathcal{C}_{#1}^{\mathrm{(eff)}}}\xspace}        
\def\Cpeff#1  {\ensuremath{\mathcal{C}_{#1}^{'\mathrm{(eff)}}}\xspace}       
\def\Ope#1    {\ensuremath{\mathcal{O}_{#1}}\xspace}                       
\def\Opep#1   {\ensuremath{\mathcal{O}_{#1}^{'}}\xspace}                    

\newcommand{\nospaceunit}[1]{\ensuremath{\text{#1}}}       
\newcommand{\aunit}[1]{\ensuremath{\text{\,#1}}}       

\newcommand{\tev}{\aunit{Te\kern -0.1em V}\xspace}
\newcommand{\gev}{\aunit{Ge\kern -0.1em V}\xspace}
\newcommand{\mev}{\aunit{Me\kern -0.1em V}\xspace}
\newcommand{\kev}{\aunit{ke\kern -0.1em V}\xspace}
\newcommand{\ev}{\aunit{e\kern -0.1em V}\xspace}
 
\newcommand{\mevc}{\ensuremath{\aunit{Me\kern -0.1em V\!/}c}\xspace}
\newcommand{\gevc}{\ensuremath{\aunit{Ge\kern -0.1em V\!/}c}\xspace}
\newcommand{\mevcc}{\ensuremath{\aunit{Me\kern -0.1em V\!/}c^2}\xspace}
\newcommand{\gevcc}{\ensuremath{\aunit{Ge\kern -0.1em V\!/}c^2}\xspace}

\def\mm   {\aunit{mm}\xspace}

\def\mum  {\ensuremath{\,\upmu\nospaceunit{m}}\xspace}

\def\fb   {\ensuremath{\aunit{fb}}\xspace}
\def\invfb   {\ensuremath{\fb^{-1}}\xspace}

\def\order{{\ensuremath{\mathcal{O}}}\xspace}
\newcommand{\chisq}{\ensuremath{\chi^2}\xspace}
\newcommand{\chisqndf}{\ensuremath{\chi^2/\mathrm{ndf}}\xspace}
\newcommand{\chisqip}{\ensuremath{\chi^2_{\text{IP}}}\xspace}

\newcommand{\chisqvs}{\ensuremath{\chi^2_{\text{VS}}}\xspace}
\newcommand{\chisqvtx}{\ensuremath{\chi^2_{\text{vtx}}}\xspace}

\def\gsim{{~\raise.15em\hbox{$>$}\kern-.85em
          \lower.35em\hbox{$\sim$}~}\xspace}
\def\lsim{{~\raise.15em\hbox{$<$}\kern-.85em
          \lower.35em\hbox{$\sim$}~}\xspace}

\def\sPlot{\mbox{\em sPlot}\xspace}

\def\pt         {\ensuremath{p_{\mathrm{T}}}\xspace}

\def\ptot       {\ensuremath{p}\xspace}

\def\evtgen     {\mbox{\textsc{EvtGen}}\xspace}

\def\geant      {\mbox{\textsc{Geant4}}\xspace}

\def\photos     {\mbox{\textsc{Photos}}\xspace}

\def\pythia     {\mbox{\textsc{Pythia}}\xspace}

\def\roofit     {\mbox{\textsc{RooFit}}\xspace}
\def\root       {\mbox{\textsc{Root}}\xspace}

\def\tell1  {TELL1\xspace}
\def\ukl1   {UKL1\xspace}

\newcommand{\eg}{\mbox{\itshape e.g.}\xspace}
\newcommand{\ie}{\mbox{\itshape i.e.}\xspace}

\newcommand{\lhcborcid}[1]{\href{https://orcid.org/#1}{\hspace*{0.1em}\raisebox{-0.45ex}{\includegraphics[width=1em]{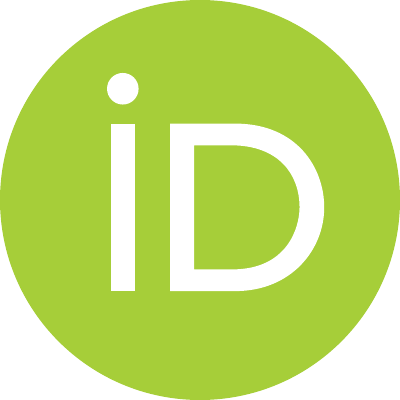}}}}

\hypersetup{
  colorlinks   = true, 
  urlcolor     = blue, 
  linkcolor    = blue, 
  citecolor    = red   
}

\ifEnableSectionTOCLinks
    \usepackage[explicit]{titlesec} 
    
    \let\oldcontentsline\contentsline
    \renewcommand

    \titleformat{\section}{\normalfont\Large\bf}{\hyperlink{tocsection.\thesection}{{\thesection} \parbox[t]{\dimexpr\textwidth-1pc}{#1}}}{1pc}{}

    \titleformat{\subsection}{\normalfont\bf}{\hyperlink{tocsubsection.\thesubsection}{{\thesubsection} \parbox[t]{\dimexpr\textwidth-1pc}{#1}}}{1pc}{}

    \titleformat{name=\section,numberless}[display]{}{}{0pt}{\normalfont\Huge\bfseries #1}
\fi

\usepackage{cite} 
\usepackage{mciteplus}

\ifthenelse{\boolean{uprightparticles}}%
{
 
}
{
 
}

\def\had  {\ensuremath{\Ph}\xspace}
\def\hadp  {\ensuremath{\Ph^+}\xspace}
\def\hadm  {\ensuremath{\Ph^-}\xspace}

\def\hadprim  {\ensuremath{\Ph^{\prime}}\xspace}

\def\hadprimm {\ensuremath{\Ph^{\prime-}}\xspace}

\def\Bdsz      {\ensuremath{\B^0_{(s)}}\xspace}

\def\BdorBdbar    {\ensuremath{\kern 0.18em\optbar{\kern -0.18em B}{}^0}\xspace}
\def\BsorBsbar    {\ensuremath{\kern 0.18em\optbar{\kern  0.06em B_s}{}^0}\xspace}

\def\pipi  {\ensuremath{\pion^+\pion^-}\xspace}

\def\BdtoKzPiPi   {\decay{\Bd}{\Kz \pip \pim}}

\def\BdtoKsKK   {\decay{\Bd}{\KS \Kp \Km}}
\def\BdtoKsPiPi   {\decay{\Bd}{\KS \pip \pim}}
\def\BdtoKsKPi   {\decay{\Bd}{\KS \Kpm \pimp}}
\def\BdtoKsKpPim   {\decay{\Bd}{\KS \Kp \pim}}
\def\BdtoKsPipKm   {\decay{\Bd}{\KS \Km \pip}}

\def\BstoKsKK   {\decay{\Bs}{\KS \Kp \Km}}
\def\BstoKsPiPi   {\decay{\Bs}{\KS \pip \pim}}
\def\BstoKsKPi   {\decay{\Bs}{\KS \Kpm \pimp}}
\def\BstoKsKpPim   {\decay{\Bs}{\KS \Kp \pim}}
\def\BstoKsPipKm   {\decay{\Bs}{\KS \Km \pip}}

\def\Kshhp{\ensuremath{\KS \hadp \hadprimm}\xspace}

\def\KsPiPi{\ensuremath{\KS \pip \pim}\xspace}
\def\KsKPi{\ensuremath{\KS \Kpm \pimp}\xspace}

\def\KsKK{\ensuremath{\KS \Kp \Km}\xspace}

\def\KsKpPim{\ensuremath{\KS \Kp \pim}\xspace}

\def\KsPipKm{\ensuremath{\KS \pip \Km}\xspace}

\def\BdstoKshhp   {\decay{\Bdsz}{\KS \hadp \hadprimm}}

\def\BdtoetapKs {\decay{\Bd}{\etapr \KS}}

\def\btoqqbars {\decay{\bquark}{\quark\quarkbar\squark}}

\def\btoccbars {\decay{\bquark}{\cquark\cquarkbar\squark}}

\def\LL   {long}
\def\DD   {downstream}

\newcommand{\Br}[1]{\ensuremath{\BF\!\left(#1\right)}\xspace}

\def\fsfdinline {\ensuremath{f_{\squark}/f_{\dquark}}\xspace}

\newcommand{\plotIfExists}[2]{
  \IfFileExists{#1}{\includegraphics[width=#2\textwidth]{#1}}{\includegraphics[width=#2\textwidth]{#1}}
}    
\newcommand{\plotOne}[2]{
  \ifthenelse{\boolean{pdflatex}}{
    \plotIfExists{#1.pdf}{#2}
  }{
    \plotIfExists{#1.eps}{#2}
  }
}

\newcommand{\plotDataFitResults}[3]{ 
  \begin{figure}[p]
    \begin{center}
      \ifthenelse{\equal{#1}{Loose}}
                 {\def \bdtComment{BDT optimisation corresponding to the favoured decay modes}}
                 {\def \bdtComment{BDT optimisation corresponding to the suppressed modes}}
                 
                 \plotOne{figs/FitResults/from5150-Louis-Polynomial-StrongPcut-#1-KSKK#2_#3-Standard-DoubleCB-NoPull_log}{0.35}
                 \plotOne{figs/FitResults/from5150-Louis-Polynomial-StrongPcut-#1-KSKK#2_#3-Standard-DoubleCB-NoPull}{0.35}\\
                 \plotOne{figs/FitResults/from5150-Louis-Polynomial-StrongPcut-#1-KSKpi#2_#3-Standard-DoubleCB-NoPull_log}{0.35}
                 \plotOne{figs/FitResults/from5150-Louis-Polynomial-StrongPcut-#1-KSKpi#2_#3-Standard-DoubleCB-NoPull}{0.35}\\
                 \plotOne{figs/FitResults/from5150-Louis-Polynomial-StrongPcut-#1-KSpiK#2_#3-Standard-DoubleCB-NoPull_log}{0.35}
                 \plotOne{figs/FitResults/from5150-Louis-Polynomial-StrongPcut-#1-KSpiK#2_#3-Standard-DoubleCB-NoPull}{0.35}\\
                 \plotOne{figs/FitResults/from5150-Louis-Polynomial-StrongPcut-#1-KSpipi#2_#3-Standard-DoubleCB-NoPull_log}{0.35}
                 \plotOne{figs/FitResults/from5150-Louis-Polynomial-StrongPcut-#1-KSpipi#2_#3-Standard-DoubleCB-NoPull}{0.35}\\
                 
                 \ifthenelse{\equal{#2}{DD}}{
                   \caption{Results of the simultaneous fit to data (\DD, #3) with the \bdtComment. The modes \KsKK, \KsKpPim, \KsPipKm and \KsPiPi are shown from top to bottom. The left-hand side plots show the results with a logarithmic scale and the right-hand side with a linear scale. Legend is similar to that of plots shown in Fig.~\ref{fig : fitLoose2011}.}
                 }{
                   \caption{Results of the simultaneous fit to data (\LL, #3) with the \bdtComment. The modes \KsKK, \KsKpPim, \KsPipKm and \KsPiPi are shown from top to bottom. The left-hand side plots show the results with a logarithmic scale and the right-hand side with a linear scale. Legend is similar to that of plots shown in Fig.~\ref{fig : fitLoose2011}.}
                 }
                 \label{fig:FitResult:#1:#2:#3}
    \end{center}
  \end{figure}
}
\newcommand{\plotSignalMCFitResults}[3]{
  \begin{figure}[!htbp]
    \begin{center}
      \plotOne{figs/FitModel/Signal/#1/Bd2KSKK#2_#3-MCFit-Louis-DoubleCB-Standard_log}{0.35}
      \plotOne{figs/FitModel/Signal/#1/Bs2KSKK#2_#3-MCFit-Louis-DoubleCB-Standard_log}{0.35}\\
      \plotOne{figs/FitModel/Signal/#1/Bd2KSKpi#2_#3-MCFit-Louis-DoubleCB-Standard_log}{0.35}
      \plotOne{figs/FitModel/Signal/#1/Bs2KSKpi#2_#3-MCFit-Louis-DoubleCB-Standard_log}{0.35}\\
      \plotOne{figs/FitModel/Signal/#1/Bd2KSpiK#2_#3-MCFit-Louis-DoubleCB-Standard_log}{0.35}
      \plotOne{figs/FitModel/Signal/#1/Bs2KSpiK#2_#3-MCFit-Louis-DoubleCB-Standard_log}{0.35}\\
      \plotOne{figs/FitModel/Signal/#1/Bd2KSpipi#2_#3-MCFit-Louis-DoubleCB-Standard_log}{0.35}
      \plotOne{figs/FitModel/Signal/#1/Bs2KSpipi#2_#3-MCFit-Louis-DoubleCB-Standard_log}{0.35}\\
      \ifthenelse{\equal{#2}{DD}}{
        \caption{Result of the simultaneous fit to simulated samples of the signal decays for \DD \KS reconstruction mode, using the \MakeLowercase{#1} optimisation of the BDT, and shown using a logarithmic scale. \KsKK, \KsKpPim, \KsPipKm, and \KsPiPi are shown from top to bottom, while \Bd decays are shown on the left and \Bs decays on the right. }
      }
                 {
                   \caption{Result of the simultaneous fit to simulated samples of the signal decays for \LL \KS reconstruction mode, using the \MakeLowercase{#1} optimisation of the BDT, and shown using a logarithmic scale. \KsKK, \KsKpPim, \KsPipKm, and \KsPiPi are shown from top to bottom, while \Bd decays are shown on the left and \Bs decays on the right. }
                 }
                 \label{fig:FitModel:Signal:#1:#2:#3}
    \end{center}
\end{figure}
}

\newcommand{\plotSplots}[3]{
  \ifthenelse{\equal{#3}{log}}
             {\def\suffix{FitStandard_log}}
             {\def\suffix{FitStandard}}
             \ifthenelse{\equal{#3}{log}}
                        {\def\scale{linear}\xspace}
                        {\def\scale{logarithmic}\xspace}
                        \begin{figure}[!htbp]
                          \begin{center}
                            \plotOne{figs/FitResults/sWeights/#1/sWeights-from5150-Louis-PolSlopes-StrongPcut-#1-KSKKDD_#2_\suffix}{0.35}
                            \plotOne{figs/FitResults/sWeights/#1/sWeights-from5150-Louis-PolSlopes-StrongPcut-#1-KSKKLL_#2_\suffix}{0.35}\\
                            \plotOne{figs/FitResults/sWeights/#1/sWeights-from5150-Louis-PolSlopes-StrongPcut-#1-KSKpiDD_#2_\suffix}{0.35}
                            \plotOne{figs/FitResults/sWeights/#1/sWeights-from5150-Louis-PolSlopes-StrongPcut-#1-KSKpiLL_#2_\suffix}{0.35}\\
                            \plotOne{figs/FitResults/sWeights/#1/sWeights-from5150-Louis-PolSlopes-StrongPcut-#1-KSpiKDD_#2_\suffix}{0.35}
                            \plotOne{figs/FitResults/sWeights/#1/sWeights-from5150-Louis-PolSlopes-StrongPcut-#1-KSpiKLL_#2_\suffix}{0.35}\\
                            \plotOne{figs/FitResults/sWeights/#1/sWeights-from5150-Louis-PolSlopes-StrongPcut-#1-KSpipiDD_#2_\suffix}{0.35}
                            \plotOne{figs/FitResults/sWeights/#1/sWeights-from5150-Louis-PolSlopes-StrongPcut-#1-KSpipiLL_#2_\suffix}{0.35}\\
                            \caption{Result of the mass fits used for the sWeights extraction with the \MakeLowercase{#1} BDT optimisation on #2 data (\scale scale). \KsKK, \KsKpPim, \KsPipKm, and \KsPiPi are shown from top to bottom, \DD on the left, \LL on the right.}
                            \label{fig:FitResult:Splots:#1:#2:#3}
                          \end{center}
                        \end{figure}
}

\newcommand{\makeSystTabular}[2]{
  \begin{table}[!htbp]
    \begin{center}
      \ifthenelse{\equal{#2}{KK}}
                 {       
                   \caption{Systematic uncertainties originating from the fit model of the \KsKK modes (using \MakeLowercase{#1} BDT optimisation). The numbers are rounded to the upper integer value, except for the total yield. The total systematic error is defined as the sum in quadrature of all the components.}
                 }{}
                 \ifthenelse{\equal{#2}{Kpi}}
                            {
                              \caption{Systematic uncertainties originating from the fit model of the \KsKpPim modes (using \MakeLowercase{#1} BDT optimisation). The numbers are rounded to the upper integer value, except for the total yield. The total systematic error is defined as the sum in quadrature of all the components.}
                            }{}
                            \ifthenelse{\equal{#2}{piK}}        
                                       {
                                         \caption{Systematic uncertainties originating from the fit model of the \KsPipKm modes (using \MakeLowercase{#1} BDT optimisation). The numbers are rounded to the upper integer value, except for the total yield. The total systematic error is defined as the sum in quadrature of all the components.}
                                       }{}
                                       \ifthenelse{\equal{#2}{pipi}}
                                                  {
                                                    \caption{Systematic uncertainties originating from the fit model of the \KsPiPi modes (using \MakeLowercase{#1} BDT optimisation). The numbers are rounded to the upper integer value, except for the total yield. The total systematic error is defined as the sum in quadrature of all the components.}
                                                  }{}
                                                  \label{Table:FitSyst:#1:#2}
                                                  \resizebox{\textwidth}{!}{
                                                    \input{tables/SystI-PolSlopes-from5150-#1-#2.txt}
                                                  }
    \end{center}
  \end{table}
  
}

\newcommand{\makeInvMass}[1]{
  \ifthenelse{\equal{#1}{Bd2KSpipi} \or \equal{#1}{Bs2KSpipi}}              {\def\myInvMass {pipi}}
             {\ifthenelse{\equal{#1}{Bd2KSpiK} \or \equal{#1}{Bs2KSpiK}}    {\def\myInvMass {piK}}
               {\ifthenelse{\equal{#1}{Bd2KSKpi} \or \equal{#1}{Bs2KSKpi}}  {\def\myInvMass {Kpi}}
                 {\ifthenelse{\equal{#1}{Bd2KSKK} \or \equal{#1}{Bs2KSKK}}  {\def\myInvMass {KK}}{\def\myInvMass ERROR}}
               }
             }

}
\newcommand{\makeMode}[1]{
  \ifthenelse{\equal{#1}{Bd2KSpipi}}                     {\def\myMode {\BdtoKsPiPi}}
             {\ifthenelse{\equal{#1}{Bd2KSpiK}}          {\def\myMode {\BdtoKsPipKm}}
               {\ifthenelse{\equal{#1}{Bd2KSKpi}}        {\def\myMode {\BdtoKsKpPim}}
                 {\ifthenelse{\equal{#1}{Bd2KSKK}}       {\def\myMode {\BdtoKsKK}}
                   {\ifthenelse{\equal{#1}{Bs2KSpipi}}           {\def\myMode {\BstoKsPiPi}}
                     {\ifthenelse{\equal{#1}{Bs2KSpiK}}          {\def\myMode {\BstoKsPipKm}}
                       {\ifthenelse{\equal{#1}{Bs2KSKpi}}        {\def\myMode {\BstoKsKpPim}}
                         {\ifthenelse{\equal{#1}{Bs2KSKK}}       {\def\myMode {\BstoKsKK}}{\def\myMode{ERROR}}
                         }
                       }
                     }
                   }
                 }
               }
             }
}

\newcommand{\plotGeomEff}[3]{
  \makeInvMass{#2}
  \makeMode{#2}
  \ifthenelse{\equal{#1}{2011}}{\def\redYear {2011}}{\def\redYear {2012}}
  \centering
  \includegraphics[width=0.45\textwidth]{figs/EfficiencyMaps/#1/Geometry/NoSel/#2/Efficiency_Map_NoSel_#2_#3_#1_\myInvMass.pdf.eps}
  \includegraphics[width=0.45\textwidth]{figs/EfficiencyMaps/#1/Geometry/NoSel/#2/Spline_Eff_NoSel_#2_#3_#1_\myInvMass.pdf.eps}
  \includegraphics[width=0.45\textwidth]{figs/EfficiencyMaps/#1/Geometry/NoSel/#2/Efficiency_errorHi_Map_NoSel_#2_#3_#1_\myInvMass.pdf.eps}
  \includegraphics[width=0.45\textwidth]{figs/EfficiencyMaps/#1/Geometry/NoSel/#2/Efficiency_errorLo_Map_NoSel_#2_#3_#1_\myInvMass.pdf.eps}
  \caption{
    (Top) $\epsilon^{\rm geom}$ as a function of the \myMode
    square Dalitz plot position obtained from \redYear-conditions generator-level signal MC:
    (left) the raw histogram,
    (right) smoothed using a 2D cubic spline.
    (Bottom) the (left) upper and (right) lower uncertainties on the histogram bins.
    Uncertainties are due to MC statistics.
  }
  \label{fig:geoeff:GeoEff:#1:#2:#3}
}

\newcommand{\plotTrackCorr}[4]{
  \makeInvMass{#2}
  \makeMode{#2}
  \centering
  \includegraphics[width=0.45\textwidth]{figs/EfficiencyMaps/#1/Tracking/#4/#2/Trk_AllEff-#4-#2-#3-#1-\myInvMass.pdf.eps}
  \includegraphics[width=0.45\textwidth]{figs/EfficiencyMaps/#1/Tracking/#4/#2/Spline_Eff_#4_#2_#3_#1_\myInvMass.pdf.eps}\\
  \includegraphics[width=0.45\textwidth]{figs/EfficiencyMaps/#1/Tracking/#4/#2/Tracking_errorHi_bootstrap_#4_#2_#3_\myInvMass_#1.pdf.eps}
  \includegraphics[width=0.45\textwidth]{figs/EfficiencyMaps/#1/Tracking/#4/#2/Tracking_errorHi_bootstrap_#4_#2_#3_\myInvMass_#1.pdf.eps}
  \caption{
    (Top) Combined tracking efficiency corrections in the \myMode signal mode for #3 and #1 configuration:
    (left) the raw histogram obtained from MC simulation and (right) smoothed using a 2D cubic spline.  
    (Bottom) the (left) upper and (right) lower uncertainties on the histogram bins.
  }
  \label{fig:DPefficiency-trk-all-#1-#2-#3-#4}
}

\newcommand{\plotTrackCorrMode}[1]{
  \begin{figure}[!htb]
    \plotTrackCorr{2011}{#1}{DD}{Loose}
  \end{figure}
  \begin{figure}[!htb]
    \plotTrackCorr{2011}{#1}{LL}{Loose}
  \end{figure}
}

\newcommand{\plotTrigCorr}[4]{
  \makeInvMass{#1}
  \makeMode{#1}
  \ifthenelse{\equal{#4}{TOS}}{
    \def\trigComment {$\epsilon^{\rm L0TOS|sel\&geom}_{\rm data}/\epsilon^{\rm L0TOS|sel\&geom}_{\rm MC}$}}{
    \def\trigComment {$\epsilon^{\rm !L0TOS|sel\&geom}_{\rm data}/\epsilon^{\rm !L0TOS|sel\&geom}_{\rm MC}$}}
  
  \centering
  \includegraphics[width=0.45\textwidth]{figs/EfficiencyMaps/2011/L0#4/#3/#1/L0#4-Correction-#3-#1-#2-2011-\myInvMass.pdf.eps}
  \includegraphics[width=0.45\textwidth]{figs/EfficiencyMaps/2011/L0#4/#3/#1/Spline_Eff_#3_#1_#2_2011_\myInvMass.pdf.eps}\\
  \includegraphics[width=0.45\textwidth]{figs/EfficiencyMaps/2011/L0#4/#3/#1/L0#4_errorHi_combined_#3_#1_#2_\myInvMass_2011.pdf.eps}
  \includegraphics[width=0.45\textwidth]{figs/EfficiencyMaps/2011/L0#4/#3/#1/L0#4_errorLo_combined_#3_#1_#2_\myInvMass_2011.pdf.eps}
  \caption{
    (Top) \trigComment across the \myMode #2 square Dalitz plot (2011+2012 combined):
    (left) the raw histogram obtained using the procedure described in the text and (right) smoothed using a 2D cubic spline.  
    (Bottom) the (left) upper and (right) lower uncertainties on the histogram bins.
  }
  \label{fig:DPcorrection-#1-#2-#3-#4}
}

\newcommand{\plotSelEff}[5]{
  \makeInvMass{#2}
  \makeMode{#2}
  \ifthenelse{\equal{#5}{TOS}}{\def\selComment {{\tt L0Hadron\_TOS}}}{\def\selComment {{\tt L0Global\_TIS\&\&!L0Hadron\_TOS}}}
  \centering
  \includegraphics[width=0.45\textwidth]{figs/EfficiencyMaps/#1/Sel#5/#4/#2/#2-Eff-#4-#2-#3-#1-\myInvMass.pdf.eps}
  \includegraphics[width=0.45\textwidth]{figs/EfficiencyMaps/#1/Sel#5/#4/#2/Spline_Eff_#4_#2_#3_#1_\myInvMass.pdf.eps}\\
  \includegraphics[width=0.45\textwidth]{figs/EfficiencyMaps/#1/Sel#5/#4/#2/#2-ErrorHi-#4-#2-#3-#1-\myInvMass.pdf.eps}
  \includegraphics[width=0.45\textwidth]{figs/EfficiencyMaps/#1/Sel#5/#4/#2/#2-ErrorLo-#4-#2-#3-#1-\myInvMass.pdf.eps}
  \caption{
    (Top) $\epsilon^{\rm sel|geom}$ across the \myMode #3 #1 square Dalitz plot for \selComment \MakeLowercase{#4} BDT candidates:
    (left) the raw histogram obtained using the procedure described in the text and (right) smoothed using a 2D cubic spline.  
    (Bottom) the (left) upper and (right) lower uncertainties on the histogram bins.
  }
  \label{fig:DPefficiency-sel-#1-#2-#3-#4-#5}
}

\newcommand{\plotSelEffMode}[2]{
  \begin{figure}[!htb]
    \plotSelEff{2011}{#1}{DD}{#1}{TOS}
  \end{figure}
  \begin{figure}[!htb]
    \plotSelEff{2011}{#1}{DD}{#1}{TIS}
  \end{figure}
  \begin{figure}[!htb]
    \plotSelEff{2011}{#1}{LL}{#1}{TOS}
  \end{figure}
  \begin{figure}[!htb]
    \plotSelEff{2011}{#1}{LL}{#1}{TIS}
  \end{figure}
}

\newcommand{\fullDataTaking}[1]
           {
             \ifthenelse{\equal{#1}{2011}}
                        {\def\fullYear {2011\xspace}}
                        {\ifthenelse{\equal{#1}{2012a}}{\def\fullYear {2012a\xspace}}{\def\fullYear {2012b\xspace}}}
           }
\newcommand{\selOptChoice}[1]
           {
	     \ifthenelse{\equal{#1}{Loose}}
	                {\def\selOpt {favoured\xspace}}
			{\def\selOpt {suppressed\xspace}}
	   }
\newcommand{\paperFitResults}[2]
           {
             \selOptChoice{#1}
               \ifthenelse{\equal{#1}{Loose}}{
               \def\myCaption{In each plot, data are the black points with error bars and the total fit model is overlaid (solid blue line).
                 The \Bd (\Bs) signal components are the pink (orange) short-dashed (dotted) lines,
                 while fully reconstructed misidentified decays are the green and dark blue dashed lines
                 close to the \Bd and \Bs peaks.
                 The sum of the partially reconstructed contributions from \B to open charm decays,
                 charmless hadronic decays, \BdtoetapKs and charmless radiative decays are
                 the red dash triple-dotted lines.
                 The combinatorial background contribution is the gray long-dash dotted
                 line.}
                 }{
                 \def\myCaption{Colours and line styles follow the same conventions as in Fig.~\ref{fig : fitLoose2011}} }
             \fullDataTaking{#2}
             \begin{figure}[!htbp]
               \begin{center}
                 \ifthenelse{\boolean{pdflatex}}{

                   \includegraphics*[width=0.49\textwidth]{figs/FusedCats-#1-KSKKDD-log.pdf}
                   \includegraphics*[width=0.49\textwidth]{figs/FusedCats-#1-KSKKLL-log.pdf}\\

                   \includegraphics*[width=0.49\textwidth]{figs/FusedCats-#1-KSKpiDD-log.pdf}
                   \includegraphics*[width=0.49\textwidth]{figs/FusedCats-#1-KSKpiLL-log.pdf}\\
                  
                   \includegraphics*[width=0.49\textwidth]{figs/FusedCats-#1-KSpipiDD-log.pdf}
                   \includegraphics*[width=0.49\textwidth]{figs/FusedCats-#1-KSpipiLL-log.pdf}

                 }{

                   \includegraphics*[width=0.49\textwidth]{figs/FusedCats-#1-KSKKDD-log.eps}
                   \includegraphics*[width=0.49\textwidth]{figs/FusedCats-#1-KSKKLL-log.eps}\\

                   \includegraphics*[width=0.49\textwidth]{figs/FusedCats-#1-KSKpiDD-log.eps}
                   \includegraphics*[width=0.49\textwidth]{figs/FusedCats-#1-KSKpiLL-log.eps}\\

                   \includegraphics*[width=0.49\textwidth]{figs/FusedCats-#1-KSpipiDD-log.eps}
                   \includegraphics*[width=0.49\textwidth]{figs/FusedCats-#1-KSpipiLL-log.eps}

                 }
               \end{center}
                 \caption{
                 Mass distributions of, from top to bottom, \KsKK, \KsKPi
                 and \KsPiPi candidates, summing the three periods of
                 data taking, with the selection optimisation chosen
                 for the \selOpt decay modes for (left) \DD~and
                 (right) \LL\ \KS reconstruction categories.
                 \myCaption
                  }                

               \label{fig : fit#1#2}
             \end{figure}
           }

\def\BodyPlotWidth {0.94}
\def\AppPlotWidth {0.95}

\begin{document}

\renewcommand{\thefootnote}{\fnsymbol{footnote}}
\setcounter{footnote}{1}


\begin{titlepage}
\pagenumbering{roman}

\vspace*{-1.5cm}
\centerline{\large EUROPEAN ORGANIZATION FOR NUCLEAR RESEARCH (CERN)}
\vspace*{1.5cm}
\noindent
\begin{tabular*}{\linewidth}{lc@{\extracolsep{\fill}}r@{\extracolsep{0pt}}}
\ifthenelse{\boolean{pdflatex}}
{\vspace*{-1.5cm}\mbox{\!\!\!\includegraphics[width=.14\textwidth]{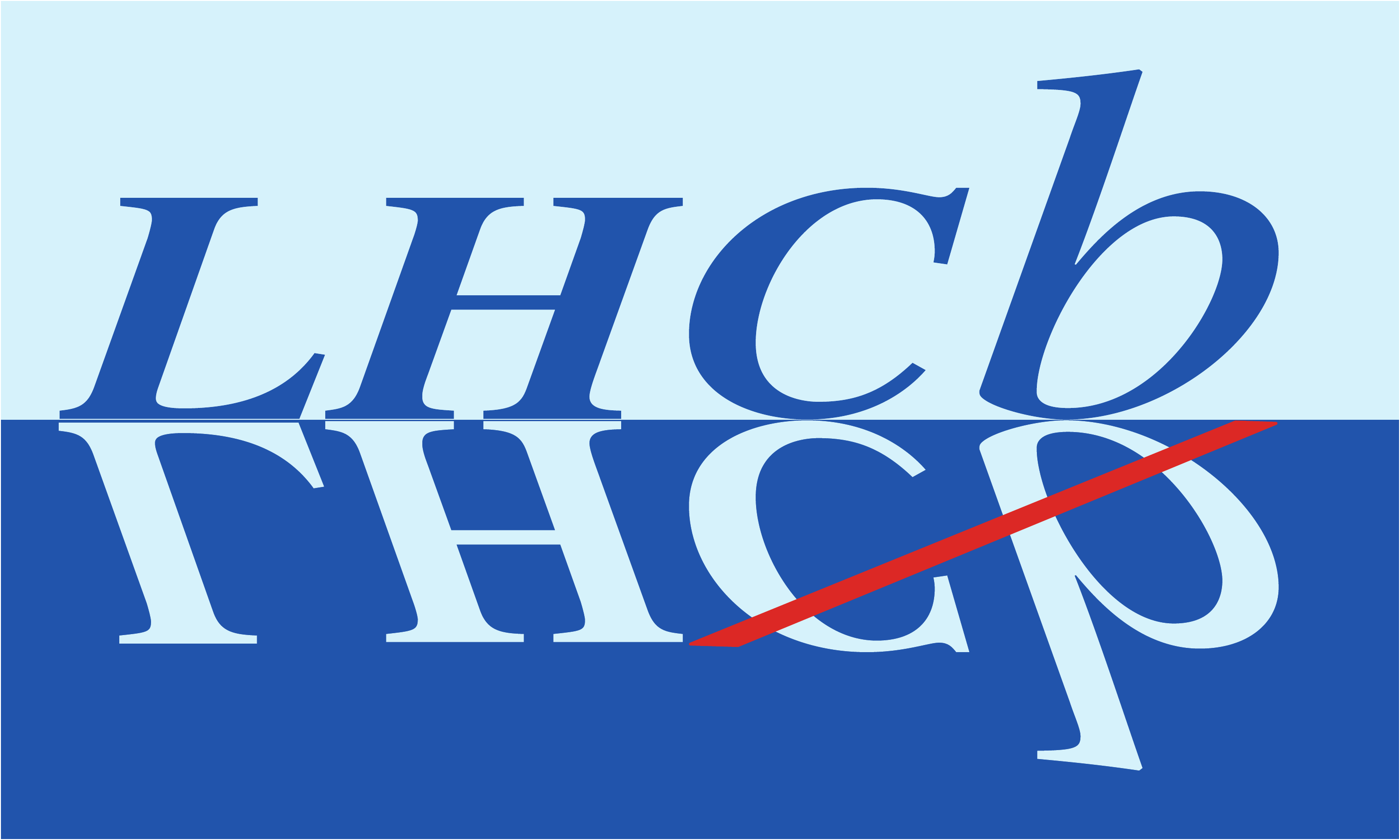}} & &}%
{\vspace*{-1.2cm}\mbox{\!\!\!\includegraphics[width=.12\textwidth]{figs/lhcb-logo.eps}} & &}%
\\
 & & CERN-EP-2026-029 \\  
 & & LHCb-PAPER-2024-029 \\  
 & & 8 July 2026 \\ 
 & & \\
\end{tabular*}

\vspace*{0.2cm}

{\normalfont\bfseries\boldmath\huge
\begin{center}
  \papertitle 
\end{center}
}

\vspace*{0.3cm}

\begin{center}
\paperauthors\footnote{Authors are listed at the end of this paper.}

\vspace*{0.3cm}

This paper is dedicated to the memory of our friend and colleague \\ Roger Barlow.

\end{center}

\vspace{\fill}

\begin{abstract}
  \noindent
  
  \noindent
This paper presents a study of the charmless three-body decays ${B^0_{(s)} \to K_{\mathrm{S}}^0 h^+ h^{\prime -}}$ (where $h^{(\prime)} = \pi, K$),
using a sample of $pp$ collision data collected by the LHCb experiment during Runs~1 and~2 of the LHC, corresponding to an integrated luminosity of $9\,\mbox{\rm fb}^{-1}$.
The  decay ${B^0_s \to K_{\mathrm{S}}^0 K^+ K^-}$ is observed for the first time,
and the following ratios of branching fractions are measured:
\begin{alignat*}{6} 
	&\frac{{\cal B}(B^0 \to K^0_{\mathrm{S}} K^+ K^-)}{{\cal B}(B^0 \to K^0_{\mathrm{S}} \pi^+\pi^-)} &&= 0.578 &&\pm 0.007 &&\pm 0.017\,, \\
	&\frac{{\cal B}(B^0 \to K^0_{\mathrm{S}}  K^\pm\pi^\mp)}{{\cal B}(B^0 \to K^0_{\mathrm{S}} \pi^+\pi^-)} &&= 0.1363 &&\pm 0.0035 &&\pm 0.0051\,, \\
	&\frac{{\cal B}(B^0_s \to K^0_{\mathrm{S}} \pi^+\pi^-)}{{\cal B}(B^0 \to K^0_{\mathrm{S}} \pi^+\pi^-)} &&= 0.269 &&\pm 0.011 &&\pm 0.015 && \pm 0.008\,, \\
	&\frac{{\cal B}(B^0_s \to K^0_{\mathrm{S}}  K^+ K^-)}{{\cal B}(B^0 \to K^0_{\mathrm{S}} \pi^+\pi^-)} &&= 0.0303 &&\pm 0.0041 &&\pm 0.0025 && \pm 0.0009\,, \\
	&\frac{{\cal B}(B^0_s \to K^0_{\mathrm{S}}  K^\pm\pi^\mp)}{{\cal B}(B^0 \to K^0_{\mathrm{S}} \pi^+\pi^-)} &&= 1.818 &&\pm 0.021 &&\pm 0.031 && \pm 0.056\,,
\end{alignat*}
  where the uncertainties are statistical, systematic, and due to knowledge of the ratio of hadronisation fractions of the $B^0_s$ and $B^0$ mesons, respectively.
\end{abstract}

\vspace*{0.3cm}

\begin{center}
  Published in Journal of High Energy Physics 07 (2026) 071
\end{center}

\vspace{\fill}

{\footnotesize 
\centerline{\copyright~\papercopyright. \href{\paperlicenceurl}{\paperlicence}.}}
\vspace*{2mm}

\end{titlepage}


\newpage
\setcounter{page}{2}
\mbox{~}


\renewcommand{\thefootnote}{\arabic{footnote}}
\setcounter{footnote}{0}


\cleardoublepage


\pagestyle{plain} 
\setcounter{page}{1}
\pagenumbering{arabic}


\section{Introduction}
\label{sec:Introduction}

The charmless three-body decay processes of neutral \B mesons, of the type \BdstoKshhp ($h^{(\prime)}=\pi,K$),\footnote{Unless stated otherwise, charge-conjugate modes are implicitly included throughout this article.} receive contributions from several Feynman-diagram topologies. In the Standard Model (SM), the decays \BdtoKsPiPi and \mbox{\BdtoKsKK} are particularly dominated by \btoqqbars ($q = u,d,s$) loop transitions. 
New particles and interactions, foreseen in various SM extensions, may alter the amplitudes responsible for the decay process~\cite{Grossman:1996ke,London:1997zk,Ciuchini:1997zp}.
Thus, measuring the branching fractions and \CP-violating asymmetries of these modes and comparing them with theoretical predictions is of great interest.
As in other multibody hadronic decay modes of \B hadrons, accurate theoretical predictions of branching fractions are challenging due to the presence of multiple interfering amplitudes.
Precision measurements of branching fractions are therefore important inputs for further refining different theoretical approaches such as QCD factorisation or perturbative QCD~\cite{Cheng:2013dua,Cheng:2014uga,Li:2014fla,Li:2014oca,Wang:2016rlo,Li:2017mao}, enabling better predictions of branching fractions and of \CP asymmetries, both for the modes considered in this paper and for other charmless decay modes.
These measurements also offer a means to evaluate the degree of isospin, U-spin, and SU(3) flavour-symmetry breaking, as illustrated, for example, in Ref.~\cite{He:2014xha}.

Studies of the decays of \Bz mesons to \KsPiPi, \KsKPi, and \KsKK final states were performed by the first-generation \B-factory experiments, \babar and Belle~\cite{Garmash:2003er,Garmash:2006fh,Aubert:2009me,delAmoSanchez:2010ur,Lees:2012kxa}.
The first study of \BdstoKshhp branching fractions performed by the \lhcb collaboration, using the data sample recorded in 2011 corresponding to an integrated luminosity of 1\invfb~\cite{LHCb-PAPER-2013-042}, reported the first observation of \BstoKsPiPi and \BstoKsKPi decays, but found no evidence for the \BstoKsKK decay.
This study was subsequently updated with the full 2011--2012 data sample, corresponding to an integrated luminosity of 3.0\invfb~\cite{LHCb-PAPER-2017-010}; once again, no conclusive evidence for the decay \BstoKsKK was found, with the significance of the corresponding signal found to be $2.5$ standard deviations.
The search reported here, using a much larger data sample, is therefore strongly motivated.

The selection criteria and mass fit procedures developed in this analysis for branching-fraction measurements may be reused in subsequent amplitude analyses, which provide access to more observables enabling detailed understanding of the decay dynamics.
In such analyses, the contributing intermediate resonances can be disentangled, allowing the branching fractions and \CP-violation parameters of the intermediate decay modes to be determined.
The measurement of such \CP-violation observables in \BdstoKshhp decays is of significant theoretical interest.
In particular, the mixing-induced \CP asymmetries in the decays dominated by loop transitions, such as \decay{\Bd}{\phi\KS}, are predicted by the Cabibbo--Kobayashi--Maskawa~(CKM) paradigm~\cite{Cabibbo:1963yz,Kobayashi:1973fv} to differ only slightly from those of \btoccbars transitions, such as \decay{\Bd}{\jpsi\KS}.
Within the SM, any deviations are expected to be small or may be controlled through flavour symmetries~\cite{Beneke:2005pu,Buchalla:2005us,Dutta:2008xw}, whereas new particles and interactions could introduce additional weak phases, potentially leading to significantly larger deviations~\cite{Grossman:1996ke,London:1997zk,Ciuchini:1997zp}.
The mixing-induced \CP-violating phase can be determined through a flavour-tagged, time-dependent amplitude analysis within the phase space of these three-body decay processes~\cite{Dalseno:2008wwa,Aubert:2009me,Nakahama:2010nj,Lees:2012kxa}.
The current experimental uncertainties~\cite{HFLAV23,PDG2024} remain substantially larger than the expected deviations, both within the SM and in scenarios extending beyond it, thereby motivating more precise measurements of these observables.
Similarly, the mixing-induced \CP-violating phase in the \Bs system could be measured through \btoqqbars processes such as \BstoKsKPi decays~\cite{Ciuchini:2007hx}.
The CKM angle \Pgamma can also be determined by combining information obtained from several \decay{\B}{\kaon\had\hadprim} decays, including the branching fractions of these modes~\cite{Ciuchini:2006kv,Gronau:2006qn,Charles:2017ptc}.
Experimental results from the \babar collaboration~\cite{BABAR:2011ae,Lees:2015uun} demonstrate the feasibility of this measurement, albeit with large statistical uncertainties.
Another method was suggested in Refs.~\cite{ReyLeLorier:2011ww,Bhattacharya:2013cla,Bhattacharya:2015uua} and applied to amplitude models previously measured by \babar~\cite{Bertholet:2018tmx}, yielding several possible values of \Pgamma.
The decay \BstoKsPiPi is dominated by tree-level processes, and thus offers a potential theoretically cleaner determination of the angle \Pgamma~\cite{Ciuchini:2006st}.

In this paper, decays of \Bd and \Bs mesons to the charmless three-body final states \KsPiPi, \KsKPi, and \KsKK are studied with the $pp$ collision data recorded by the LHCb detector,
corresponding to an integrated luminosity of 1.0\invfb at a centre-of-mass energy of $\sqrt{s}=7\tev$ in 2011, 2.0\invfb at $\sqrt{s}=8\tev$ in 2012, and 6\invfb at $\sqrt{s}=13\tev$ from 2015 to 2018.
The integrated luminosity is approximately three times larger than that used in Ref.~\cite{LHCb-PAPER-2017-010}, which this paper supersedes.
Furthermore, the production cross-section of \Bd and \Bs mesons increases with the centre-of-mass energy~\cite{LHCb-PAPER-2016-031}.
The time-integrated, \CP-averaged branching fractions are measured relative to the \BdtoKsPiPi decay.
For \Bs decays, this implies that the contribution arising from nonzero width difference is included~\cite{DeBruyn:2012wj}.
Throughout this document, the notation \Br{\Bz\to\KS\Kpm\pimp} refers to the sum of the branching fractions \Br{\Bz\to\KS\Kp\pim} and \Br{\Bz\to\KS\Km\pip}, and the same applies to the corresponding \Bs decays.

\section{Detector and simulation}
\label{sec:Detector}

The \lhcb detector~\cite{LHCb-DP-2008-001,LHCb-DP-2014-002} is a single-arm forward 
spectrometer covering the \mbox{pseudorapidity} range $2<\eta <5$,
designed for the study of particles containing \bquark or \cquark
quarks. The detector  used to collect the data analysed in this paper includes a high-precision tracking system
consisting of a silicon-strip vertex detector surrounding the $pp$
interaction region (the \velo), a large-area silicon-strip detector located
upstream of a dipole magnet with a bending power of about
$4{\mathrm{\,T\,m}}$, and three stations of silicon-strip detectors and straw
drift tubes
placed downstream of the magnet.
The tracking system provides a measurement of the momentum, \ptot, of charged particles with
a relative uncertainty that varies from 0.5\% at low momentum to 1.0\% at 200\gevc.
The minimum distance of a track to a primary $pp$ collision vertex (PV), the impact parameter (IP),
is measured with a resolution of $(15+29/\pt)\mum$,
where \pt is the component of the momentum transverse to the beam, in\,\gevc.
Different types of charged hadrons are distinguished using information
from two ring-imaging Cherenkov detectors.
Photons, electrons and hadrons are identified by a calorimeter system consisting of
scintillating-pad and preshower detectors, an electromagnetic
and a hadronic calorimeter. Muons are identified by a
system composed of alternating layers of iron and multiwire
proportional chambers.

Simulation is required to model the effects of the detector acceptance and selection requirements and to investigate backgrounds from other $b$-hadron decays.
  In the simulation, $pp$ collisions are generated using
  \pythia~\cite{Sjostrand:2007gs,*Sjostrand:2006za}
  with a specific \lhcb configuration~\cite{LHCb-PROC-2010-056}.
  Decays of unstable particles
  are described by \evtgen~\cite{Lange:2001uf}, in which final-state
  radiation is generated using \photos~\cite{davidson2015photos}.
  The interaction of the generated particles with the detector, and its response,
  are implemented using the \geant
  toolkit~\cite{Allison:2006ve, *Agostinelli:2002hh} as described in
  Ref.~\cite{LHCb-PROC-2011-006}.

To account for possible mismodelling of the detector response in simulation, in particular that of the particle identification (PID) systems, dedicated data and simulated calibration samples are used to correct the values of the corresponding variables for each candidate reconstructed in the simulated signal samples.
The PID corrections are based on matching the cumulative distribution function (CDF) of the variable in question between the data and simulation calibration samples~\cite{LHCb-PUB-2016-021}.
The variation of the PID response with the kinematics of the track and the occupancy of the detector is accounted for using a kernel density estimation in the four dimensions: PID variable CDF, \pt and $\eta$ of the track, and the track multiplicity of the event~\cite{Poluektov:2014rxa}.

\section{Trigger and event selection}
\label{sec:Selection}

The online event selection is performed by a trigger~\cite{LHCb-DP-2012-004}, which consists of a hardware stage, based on information from the calorimeter and muon systems, followed by a software stage, which applies a full event reconstruction.
At the hardware trigger level, events are required to have a muon with high \pt or a hadron, photon or electron with high transverse energy in the calorimeters.
For the decay modes considered, all of these triggers can be caused by activity in the event not associated with the signal candidate and, for the hadron triggers, they can also be caused by the signal candidate itself.
The software trigger requires a two-, three- or four-track secondary vertex with a significant displacement from all primary $pp$ interaction vertices.
At least one charged particle must have high transverse momentum
and be inconsistent with originating from any reconstructed PV.
A multivariate algorithm~\cite{BBDT,LHCb-PROC-2015-018} is used for the identification of secondary vertices consistent with the decay of a \bquark hadron.
It is required that the software trigger decision must have been caused entirely by tracks from the decay of the signal \B candidate.

To suppress background, events satisfying the trigger requirements are filtered in two subsequent stages: a set of loose preselection requirements, followed by an optimised multivariate selection.
A potential source of systematic uncertainty is the accuracy of modelling the variation of the selection efficiency over the phase space of the \B decay.
To mitigate this, the selection is designed to minimise efficiency variations by imposing only loose requirements on properties (such as the momenta) of the individual \B-meson decay products.
Instead, extensive use is made of topological features such as the flight distance of the \B candidate.
The calculated values of these features depend on the assumption that the \B candidate (or its decay products) have originated from a particular PV.
To remove ambiguity, it is therefore necessary to associate each candidate with a single PV, that from which it is most consistent with having originated.
The metric for defining the consistency is the \chisqip quantity, which is the difference in vertex-fit \chisq of the given PV reconstructed with and without the candidate in question.
In events that contain more than one PV, each candidate is associated with the PV with which it forms the smallest \chisqip value.

Candidate \decay{\KS}{\pip\pim} decays are reconstructed in two different categories:
the first involving \KS mesons that decay early enough for the resulting pions to be reconstructed in the \velo;
and the second contains those \KS mesons that decay further downstream, such that track segments of the pions cannot be formed in the \velo.
These \KS reconstruction categories are referred to as \emph{\LL}~(LL) and \emph{\DD}~(DD), respectively.
The \LL\ category has better mass, momentum and vertex resolution than the \DD\ category.
There are, however, approximately twice as many \KS candidates reconstructed as DD than as LL, due to the lifetime of the \KS meson and the geometry of the detector.
In the following, \B candidates whose decay products include either a \LL\ or \DD\ \KS candidate are referred to using the corresponding category names.
During the 2012 data taking, a significant improvement of the trigger efficiency for \DD\ \KS candidates was obtained following an update of the software trigger algorithms.
To take this into account, the 2012 data sample is subdivided into the 2012a and 2012b data-taking periods.

The \KS candidates are formed by combining two oppositely charged pions, both of which are required to have momentum $\ptot>2\gevc$ and have \chisqip with respect to their associated PV greater than 9 (4) for \LL\ (\DD) candidates.
The vertex of the \KS candidate is determined
and the resulting mass must be within $\pm20$\mevcc ($\pm30$\mevcc) of the known \KS mass~\cite{PDG2024} for \LL\ (\DD) candidates.
To discriminate against tracks originating from the PV, the \KS vertex is required to be significantly displaced from its associated PV, as quantified by $\chisqvs$, the difference in vertex-fit $\chi^2$ when the secondary vertex is constrained to coincide with the primary vertex.
In addition, \DD\ \KS candidates are required to have a momentum $\ptot>6\gevc$.

The \B candidates are formed by combining a \KS candidate and two oppositely charged particles, $\hadp\hadprimm$, initially assuming the pion-mass hypothesis.
The momenta of both charged particles are required to be in the range $3<\ptot<100\gevc$ to ensure good pion-kaon discrimination.
Both charged particles should also be inconsistent with being muons based on information from the muon system.
The scalar sum of the transverse momenta of the \KS, \hadp, and \hadprimm candidates must be greater than 3.0\gevc (4.2\gevc), for \LL\ (\DD) candidates, and at least two of the three decay products must have $\pt>0.8\gevc$.
The \B candidates must have $\pt>1.5\gevc$ and mass within the range $4000 < m_{\KsPiPi} < 6200 \mevcc$.
The fitted \B-candidate decay vertex should be
significantly separated from any PV,
and isolated from other tracks, as quantified by
the change in vertex-fit chi-square, \chisqvtx, when adding any other track in the event.
The \B candidates should also be consistent with originating from a PV, quantified by requiring, for \LL\ (\DD) candidates, both that $\chisqip < 8 \, (6)$ and that the cosine of the angle $\theta_{\rm DIR}$ between the reconstructed momentum vector of the \B candidate and the vector between the associated PV and the decay vertex be greater than 0.9999 (0.999).
Finally, the decay vertex of the \KS candidate is required to be at least 30\mm downstream, along the beam direction, from that of the \B candidate.
In conjunction with the \KS mass requirement, this suppresses direct four-body decays \decay{B^0_{(s)}}{h^+ h^{\prime -} \pi^+ \pi^-}, as well as misreconstructed \decay{B_{(s)}^0}{\KS h^+ h^{\prime-}} decays in which a pion produced in the \KS decay is exchanged with the $h$ or $h^\prime$ track of the same charge.

The next stage of the selection uses dedicated multivariate discriminants trained using the \textsc{XGBoost} implementation~\cite{XGBoost} of a boosted decision tree~(BDT) algorithm~\cite{Breiman}.
Two different types of classifiers are trained, each designed to suppress a particular background contribution.
The first is designed to suppress the combinatorial background and mainly uses information on the decay topology.
Since the different signal channels have almost identical topologies, the same BDT is used for all channels.
In contrast, the second classifier, referred to as the PID BDT in the following, is designed to suppress background from so-called cross-feed backgrounds, where a signal decay from one signal channel is misidentified as being from another signal channel, \eg a \BstoKsKPi decay where the final-state kaon is misidentified as a pion and so is reconstructed as a candidate in the \KsPiPi spectrum.
Accordingly, this latter classifier is trained using information from the particle-identification systems and separate classifiers are trained for each of the signal final states.

The features used in the topological BDT training are:
the \pt, $\eta$, \chisqip, \chisqvs, $\cos\theta_{\rm DIR}$ and \chisqvtx values of the \B candidate;
the smallest change in the \B-candidate \chisqvtx value when adding another track from the event;
the sum of the \chisqip values of the \hadp and \hadprimm candidates and, in the case of the \LL\ category, additionally that of the \KS candidate;
and the \chisqvs of the \KS candidate for the \LL~category only.
Additional variables included are four asymmetries of the form
\begin{equation}
X^\mathrm{asym} \equiv \frac{X^P - X^\mathrm{cone}}{X^P + X^\mathrm{cone}} \,,
\end{equation}
where $X^P$ is the quantity $X$ (\ptot or \pt) for particle $P$ (\B or \KS)
and $X^\mathrm{cone}$ is the same quantity determined from a vector sum of the momenta of all charged tracks inside a cone around the direction of particle $P$,
of radius $R \equiv \sqrt{\delta\eta^2 + \delta\phi^2} = 1.5$.
Here $\delta\eta$ and $\delta\phi$ are the difference in pseudorapidity and azimuthal angle (in radians) around the beam direction,
between the momentum vector of the track under consideration and that of particle $P$.
Finally, four differences of the form
\begin{equation}
Y^\mathrm{diff} \equiv Y^P - Y^\mathrm{cone} \,,
\end{equation}
where the quantity $Y$ is either $\eta$ or $\phi$, are also included in the BDT.
The signal and background training samples are, respectively, simulated \BdtoKsPiPi decays and data from the upper mass sideband, $5425 < m_{\KsPiPi} < 6200 \mevcc$.
Contributions from \decay{\Lb}{\Lc\hadm}, \decay{\Lc}{\KS\proton} decays are removed from the training samples taken from data, using mass vetoes.

The features used to train the PID BDT are the outputs of multivariate classifiers applied to the \hadp and \hadprimm candidates, each designed to identify kaons, pions, and protons using information from the particle-identification and tracking detectors.
The PID BDT for a given final state is trained using simulation, where the signal training sample consists of decays of the channel in question, while the background training sample contains different \BdstoKshhp decay modes.
For both signal and background training samples, the candidates are reconstructed under the mass hypothesis of the final state being considered, meaning that in the background sample one of the final-state hadrons has been assigned the wrong identity.

The selection requirements placed on the output of the two BDTs are simultaneously optimised.
The optimisation is performed independently for each data-taking year, with 2012 subdivided as mentioned above, and for each \KS reconstruction type.
It is also performed separately for the \Bd and \Bs signals that are present in each final state, since the expected yields of the two decays can differ greatly.
The selection optimisation for those modes that are favoured (\BdtoKsPiPi, \BstoKsKPi, \BdtoKsKK) is referred to as looser, and the selection optimisation for those modes that are suppressed (\BstoKsPiPi, \BdtoKsKPi, \BstoKsKK) is referred to as tighter.
For all signal decay modes that have previously been observed, the following figure of merit is used
\begin{equation}
{\cal Q}_1 \equiv \frac{N_{\rm sig}}{\sqrt{N_{\rm sig}+N_{\rm bg}}} \,,
\end{equation}
where $N_{\rm sig}$ ($N_{\rm bg}$) represents the number of expected signal (combinatorial background) candidates that are retained by a given selection and fall within the signal region.
The signal region is defined as a mass window of $\pm 2.5$ times the estimated signal resolution around the known \Bd or \Bs mass~\cite{PDG2024}, as appropriate.
The value of $N_{\rm sig}$ is estimated based on the known branching fractions~\cite{PDG2024} and the signal efficiency ($\eps_{\rm sig}$) determined from simulation.
Due to the simultaneous optimisation of the two BDTs, the calculation of $N_{\rm bg}$ must include both combinatorial and cross-feed background sources.
The expected number of cross-feed candidates is calculated similarly to $N_{\mathrm{sig}}$,
while the number of expected combinatorial candidates is determined by fitting the mass distribution above the signal region ($5550 < m_{\Kshhp} < 6200 \mevcc$) and extrapolating into the signal region.
The \BstoKsKK mode is yet to be observed and so an alternative figure of merit is used that does not require knowledge of the signal branching fraction~\cite{Punzi:2003bu},
\begin{equation}
	{\cal Q}_2 \equiv \frac{\eps_{\rm sig}}{\frac{5}{2} + \sqrt{N_{\rm bg}}} \,,
\end{equation}
where the value of $N_{\rm bg}$ is estimated in the same way as described for $\mathcal{Q}_1$.

A final category of background that must be considered consists of fully reconstructed \bquark-hadron decays into a \Kshhp final state that proceed via an intermediate charm or charmonium state,
\eg two-body $\D\had$, $\Lc\had$ or $(\cquark\cquarkbar)\KS$ combinations, where $(\cquark\cquarkbar)$ indicates a charmonium resonance.
Such decays often satisfy the selection criteria with reasonably high efficiency and may result in a very similar \B-candidate mass distribution as the signal candidates.
Therefore, the following decays are explicitly reconstructed under the relevant particle mass hypotheses and vetoed in all the spectra:
\decay{\Dz}{\Km\pip},
\decay{\Dp}{\KS\Kp},
\decay{\Dp}{\KS\pip},
\decay{\Dsp}{\KS\Kp},
\decay{\Dsp}{\KS\pip},
\decay{\Lc}{\proton\KS},
\decay{\jpsi}{\pipi},
\decay{\jpsi}{\Kp\Km},
\decay{\chiczero}{\pipi},
and
\decay{\chiczero}{\Kp\Km}.
The vetoed mass region for each reconstructed charm (charmonium) state is a window of $\pm 30\,(\pm48)\mevcc$ around the known mass of that state~\cite{PDG2024}.
This range reflects the typical mass resolution obtained at \lhcb.
Similarly, backgrounds from \decay{\Lz}{\proton\pim} decays being reconstructed as \decay{\KS}{\pip\pim} are suppressed by applying a mass veto window of $\pm 6.8\mevcc$ around the known \Lz mass~\cite{PDG2024} when applying the $\proton\pim$ or $\pip\antiproton$ mass hypothesis to the \KS-candidate decay products.

\section{Fit model}
\label{sec:FitModel}

The data are divided into seven statistically independent subsamples, separated by data-taking period (2011, 2012a, 2012b, 2015, 2016, 2017, 2018).
The selected $B$ candidates form eight mass spectra: four final states ($\KS \pip \pim$, $\KS \Kp \pim$, $\KS \pip \Km$, $\KS \Kp \Km$) multiplied by two \KS reconstruction types (LL, DD).
For each subsample,
simultaneous unbinned maximum-likelihood fits to the eight mass spectra in the range 5100--5750\mevcc are used to determine the signal yields.
In general, each final state will contain two signal peaks, \Bz and \Bs, and, as discussed above, the optimum working point in the selection is significantly different between the two peaks (referred to as the tighter and looser selections for the suppressed and favoured decay modes, respectively). To minimise the expected statistical uncertainties on the ratios of branching fractions, the fit procedure described below is performed twice for each subsample: once with the tighter selections, to be used to measure the branching fractions of the \decay{\Bd}{\KS K^{\pm} \pi^{\mp}}, \decay{\Bs}{\KS \pip \pim}, and \decay{\Bs}{\KS \Kp \Km} decays relative to that of the \decay{\Bz}{\KS \pip \pim} decay; and once with the looser selections, to be used to measure the branching fractions of the \decay{\Bz}{\KS \Kp \Km} and \decay{\Bs}{\KS K^{\pm} \pi^{\mp}} decays relative to that of the \decay{\Bz}{\KS \pip \pim} decay.

Several different components contribute to each mass spectrum. Those considered in the baseline fit model are
  the two signal peaks arising from correctly reconstructed \Bd and \Bs signal decays,
  cross-feed structures due to other \decay{B_{(s)}^0}{\KS h^+ h^{\prime -}} decays in which one or both charged hadrons is misidentified,
  partially reconstructed decays of a $b$~hadron to a final state of the form $\KS h^+ h^{\prime -} X$ where the system $X$ is not reconstructed, and
  combinatorial background.

Each signal peak is modelled as the sum of two Crystal Ball functions~\cite{Skwarnicki:1986xj}, which share common peak position ($\mu$) and width ($\sigma$) parameters, but whose power-law tails are on opposite sides of the peak and have separate parameters ($n_L$, $n_R$, $\alpha_L$, $\alpha_R$).
An additional parameter ($f_L$) determines the fraction of candidates in the left-tailed Crystal Ball function, for a total of seven shape parameters.
A simultaneous fit to the eight mass spectra in simulated signal events of the subsample under consideration is used to determine the lineshape parameters, with certain parameters (or ratios of parameters) shared between mass spectra and between the \Bz and \Bs peaks; see Appendix~\ref{sec:app-lineshapes} for details. In particular, the peak position parameters $\mu(\Bz)$ and $\mu(\Bs)$ are common across all spectra, and the width parameters $\sigma$ are defined relative to one reference peak, corresponding to the \decay{\Bz}{\KS \pip \pim} decay with \DD\ \KS reconstruction. When fitting to a subsample of data, only two of the signal shape parameters are free to vary, with all others being derived from those two or fully fixed from the fit to simulated data. The two free shape parameters are the \Bz peak position parameter, with the \Bs peak position derived according to $\mu(\Bs) = \mu(\Bz) + \Delta m$, where $\Delta m = 87.26$\mevcc is the known difference in mass between the two states~\cite{PDG2024}; and the peak width parameter $\sigma(\Bz \to \KS \pip \pim ; \mathrm{DD})$, with all other width parameters derived from it according to the ratios seen in the fit to simulation.
It has been determined that the ratio of the widths in the \DD\ and \LL\ \KS reconstruction types of the \BdtoKsPiPi channel is well reproduced in the simulation.
All signal yields are free to vary in the fit to data.

Cross-feed components occur due to hadron misidentification of the $h^+$ or $h^{\prime -}$ mesons, such that events from one of the eight signal decay modes appear in a different mass spectrum, \eg \decay{\Bz}{\KS \pip \pim} decays may appear in the $\KS \Kp \pim$ spectrum if the \pip is misidentified as a \Kp.
The probability for a misidentified decay to pass the selection is determined with simulation, corrected for known data-simulation differences, in the same way as the efficiency for a correctly identified decay to pass the selection, as described in Sec.~\ref{sec:Systematics}.
Cross-feed components are modelled similarly to the signal, using the sum of two Crystal Ball functions. For each combination of a true decay and reconstructed final state, the shape parameters are determined from a fit to simulated signal events of the true decay that pass the selection for the reconstructed final state. This fit is performed simultaneously to all data-taking periods and to both \KS reconstruction types, and with the final states $\KS \Kp \pim$ and $\KS \pip \Km$ combined.
Double misidentification (\eg \decay{\Bz}{\KS \pip \pim} decays identified as $\KS \Kp \Km$) is found to be negligible, as is cross-feed from the suppressed decays \decay{\Bs}{\KS \pip \pim} and \decay{\Bs}{\KS \Kp \Km}.
This leaves two cross-feed structures in each final state:
\decay{\Bz}{\KS K^\pm \pi^\mp} and \decay{\Bs}{\KS K^\pm \pi^\mp} in $\KS \pip \pim$;
 \decay{\Bz}{\KS \pip \pim} and \decay{\Bz}{\KS \Kp \Km} in $\KS K^\pm \pi^\mp$;
 and \decay{\Bz}{\KS K^\pm \pi^\mp} and \decay{\Bs}{\KS K^\pm \pi^\mp} in $\KS \Kp \Km$.
When fitting to a subsample of data, the shape of each cross-feed component is fixed from the fit to simulated data.
Since the misidentification probabilities are known from simulation within some uncertainty, 
the yield of each misidentified cross-feed contribution may be appropriately constrained from this information and the yield of the corresponding correctly identified signal component with the same \KS reconstruction.
For each final state, this is implemented by constraining the ratio of the yields of the two cross-feed structures; this leaves one free cross-feed yield-normalisation parameter per final state.
In addition, when applying the tighter selection, a second constraint is added, such that the cross-feed yields are fully determined by the correctly identified signal yields and the misidentification probabilities.
The uncertainty arising from the precision and accuracy of the misidentification probabilities derived from simulation is propagated as part of the systematic uncertainties.

A range of different physical processes contribute to partially reconstructed components;
  in each case, one or more final-state particles from the true decay are omitted in the reconstruction
  (\eg \decay{\Bz}{K^{*0} \pi^+ \pi^-}, \decay{K^{*0}}{\KS \pi^0} with the $\piz$ omitted).
Because part of the four-momentum is missing in such decays, the lineshape is, in general, displaced to lower mass.
Since the fit range starts at 5100\mevcc, a complete description of all partially reconstructed decays is not required: often, the bulk of their distribution lies below the lower fit boundary. Instead, simulated samples of a small number of decay modes are used,
and in each case the lineshape is modelled as the convolution of a generalised ARGUS function~\cite{Albrecht:1990cs} with a Gaussian function,
using a simultaneous fit to all data-taking periods and to both \KS reconstruction types.
For partially reconstructed charmless decays, the representative decay mode used is
\decay{\Bz}{K^{*}(892)^0 \rho(770)^0}, with \decay{K^{*}(892)^0}{\KS \piz} and \decay{\rho(770)^0}{\pip \pim}
with the \piz omitted,
and the lineshape obtained is used for all final states and for both partially reconstructed $B_{d/u}$ and \Bs decays (with an offset of $+\Delta m$ when the source is a \Bs decay).
When fitting to a subsample of data, the shape of each partially reconstructed component is fixed from the fits to simulated data.
Each final state contains two charmless partially reconstructed components ($B_{d/u}$ and \Bs), and their yields are free to vary in the fit to data. Other contributions are considered as systematic variations, as described in Sec.~\ref{sec:Systematics-fit}.

The combinatorial background component, arising from unrelated or uncorrelated combinations of final-state particles, is parameterised as a first-order polynomial function.
When fitting to a subsample of data, the yield and slope of the combinatorial component are free to vary for each final state, except that the $\KS \Kp \pim$ and $\KS \pip \Km$ final states share a common slope parameter.

\begin{figure}[!htb]
  \begin{center}
  \includegraphics[width=\BodyPlotWidth\textwidth]{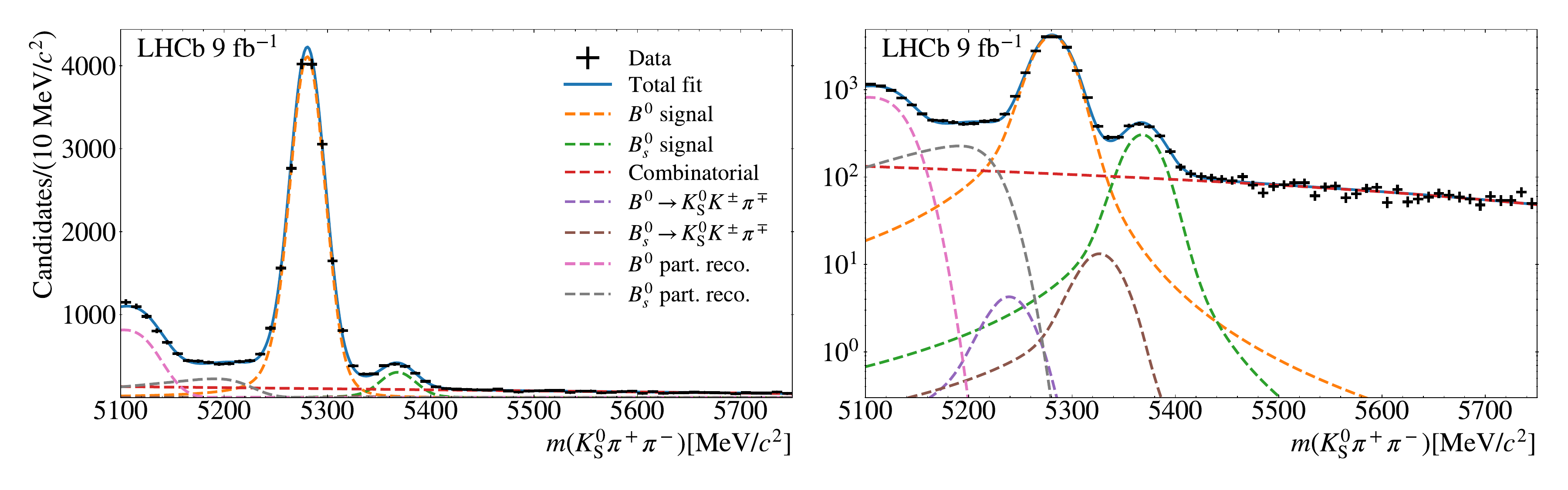}
  \includegraphics[width=\BodyPlotWidth\textwidth]{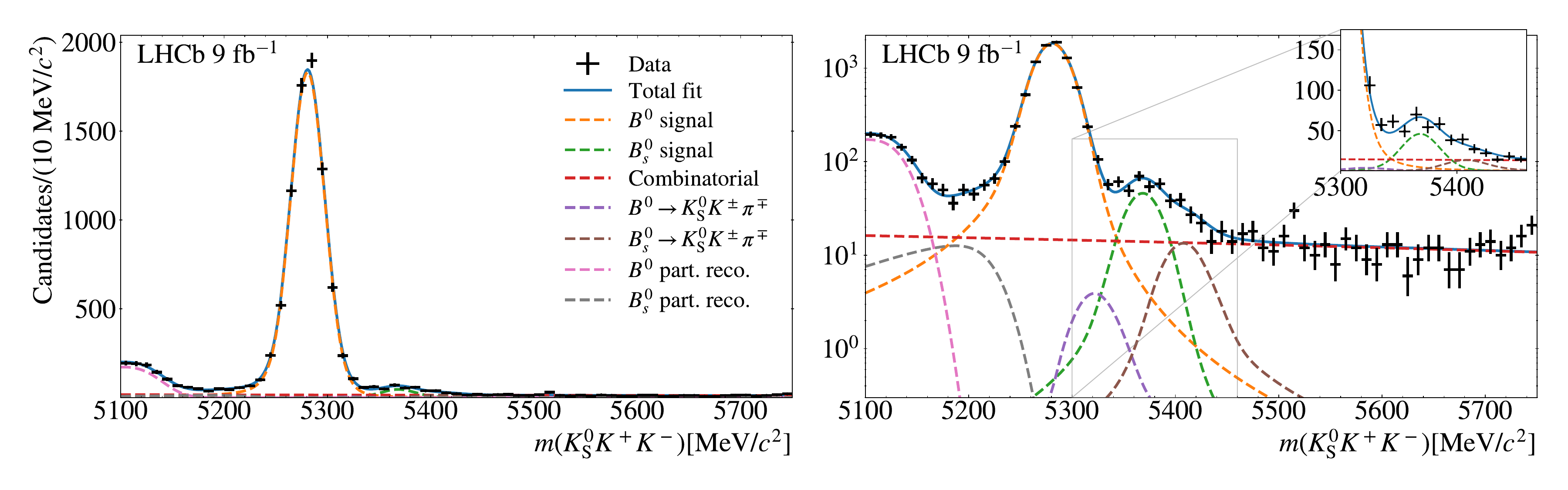}
  \includegraphics[width=\BodyPlotWidth\textwidth]{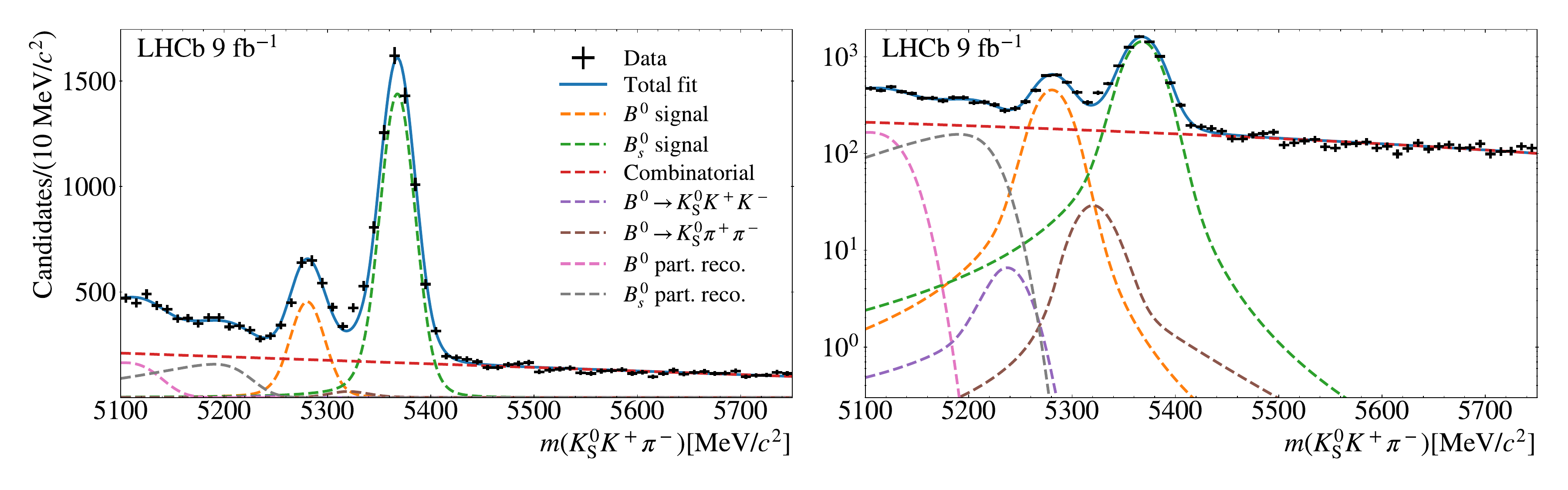}
  \includegraphics[width=\BodyPlotWidth\textwidth]{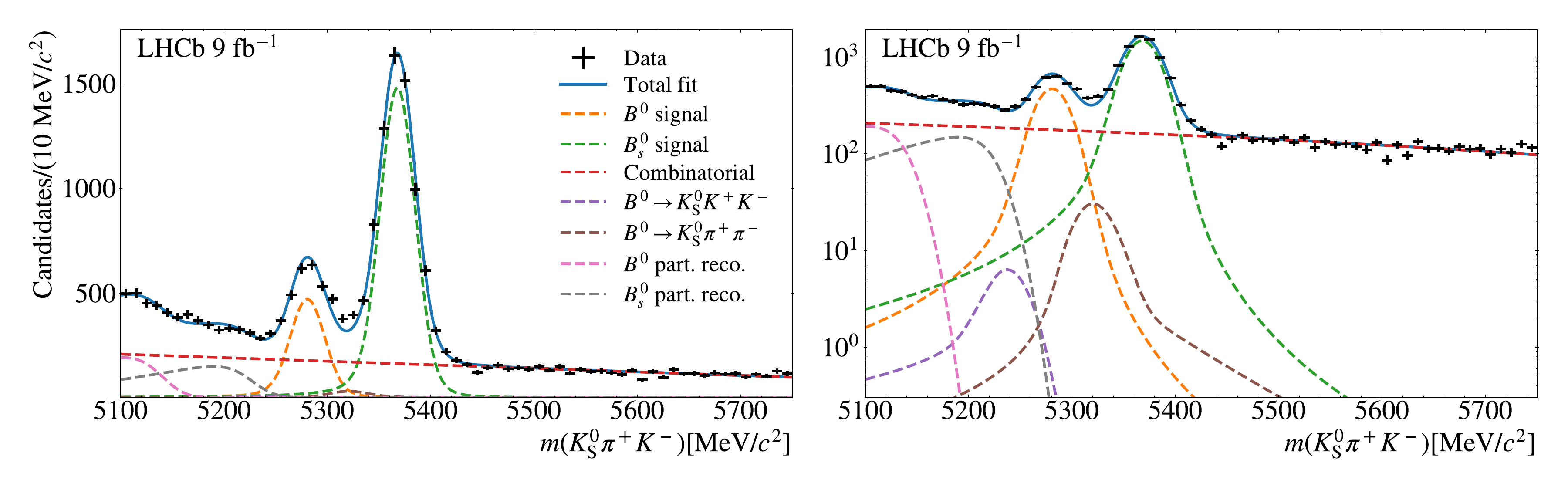}
  \end{center}
  \caption{
    \small 
      Mass distributions of $K^0_{\mathrm{S}} h^+ h^{\prime -}$ candidates that pass the tighter selection, combining all running periods and both $K^0_{\mathrm{S}}$ reconstruction types.
      The tighter selection is used to measure the ratios of branching fractions with the following decays in the numerator: $B^0_s \to K^0_\mathrm{S} \pi^+ \pi^-$, $B^0 \to K^0_\mathrm{S} K^\pm \pi^\mp$, $B^0_s \to K^0_\mathrm{S} K^+ K^-$.
      For illustration, the sum of the fits to the subsamples is also shown.
      From top to bottom, each row corresponds to a different final state:
      $K^0_{\mathrm{S}} \pi^+ \pi^-$,
      $K^0_{\mathrm{S}} K^+ K^-$,
      $K^0_{\mathrm{S}} K^+ \pi^-$,
      $K^0_{\mathrm{S}} \pi^+ K^-$.
      The inset zooms into the ${B^0_s \to K^0_{\mathrm{S}} K^+ K^-}$ signal region.
       }
  \label{fig:fit-sum-SP}
\end{figure}

\begin{figure}[!htb]
  \begin{center}
  \includegraphics[width=\BodyPlotWidth\textwidth]{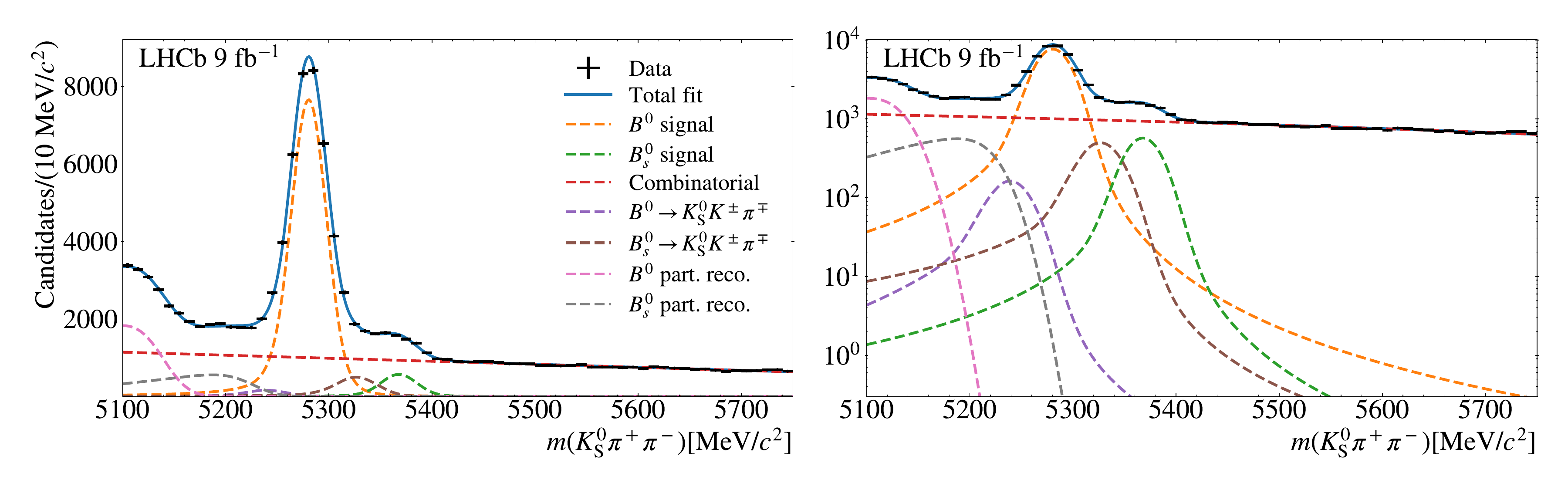}
  \includegraphics[width=\BodyPlotWidth\textwidth]{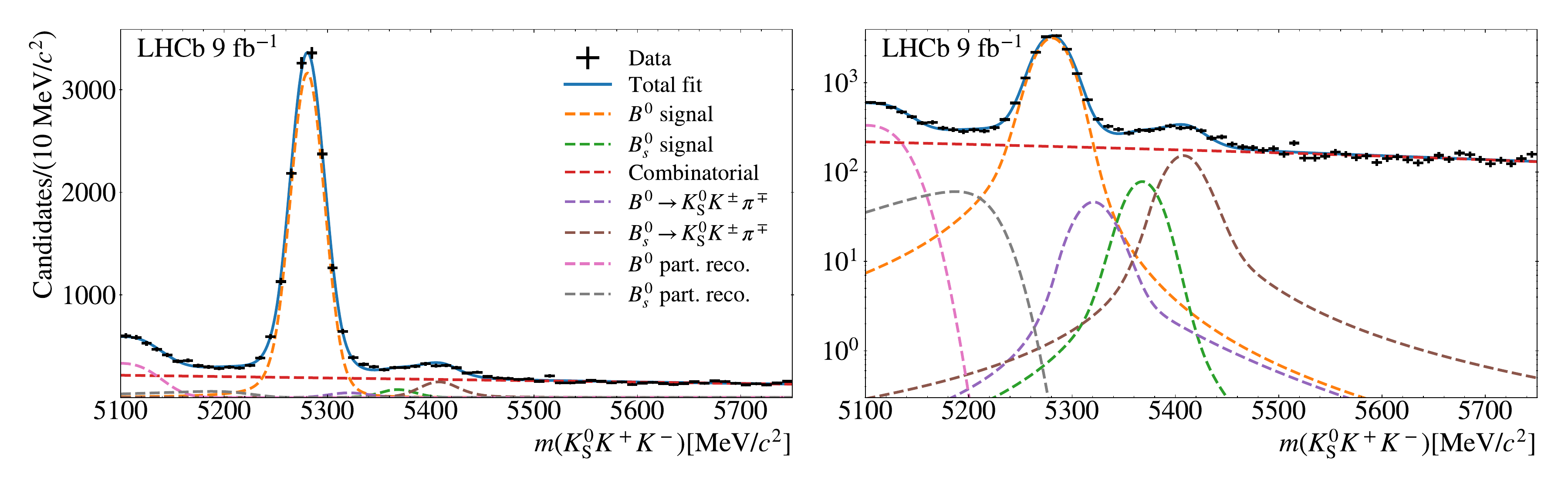}
  \includegraphics[width=\BodyPlotWidth\textwidth]{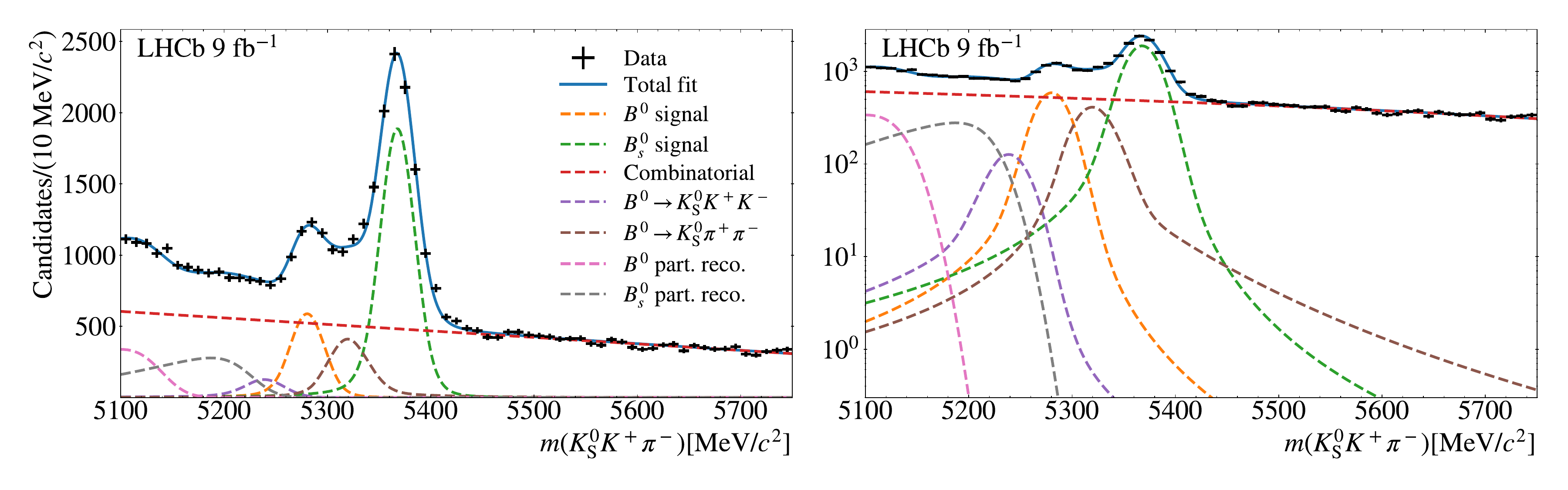}
  \includegraphics[width=\BodyPlotWidth\textwidth]{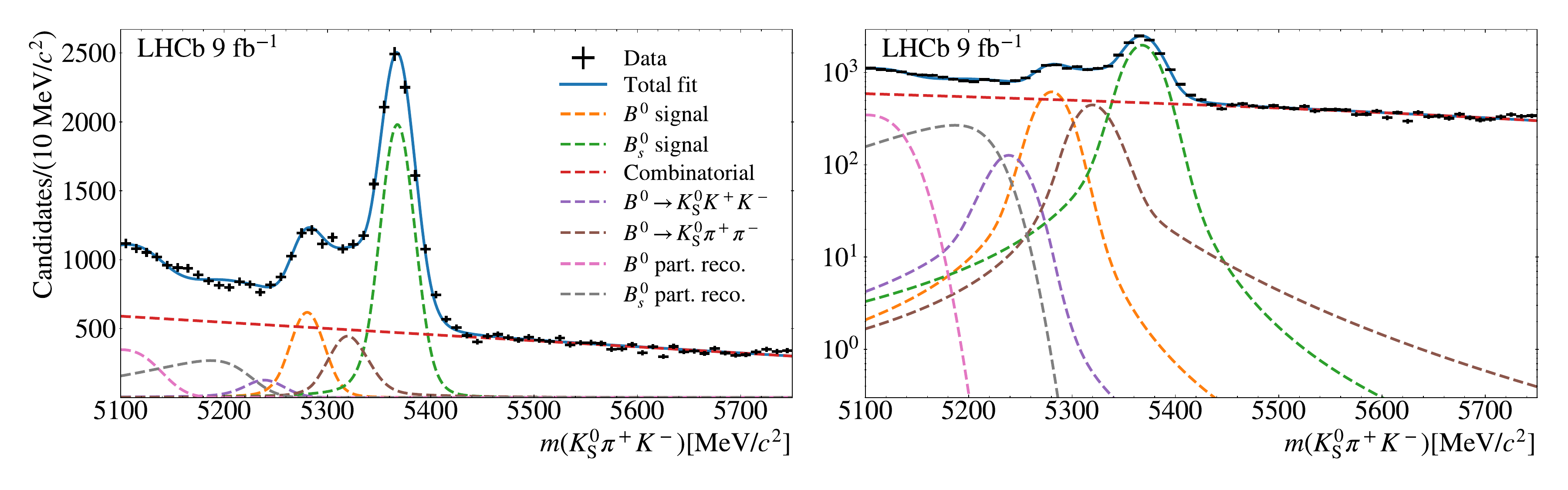}
  \end{center}
  \caption{
    \small 
      Mass distributions of $K^0_{\mathrm{S}}  h^+ h^{\prime -}$ candidates that pass the looser selection, combining all running periods and both $K^0_{\mathrm{S}}$ reconstruction types.
      The looser selection is used to measure the ratios of branching fractions with the following decays in the numerator: $B^0_s \to K^0_\mathrm{S} K^\pm \pi^\mp$, $B^0 \to K^0_\mathrm{S} K^+ K^-$.
      For illustration, the sum of the fits to the subsamples is also shown.
      From top to bottom, each row corresponds to a different final state:
      $K^0_{\mathrm{S}} \pi^+ \pi^-$,
      $K^0_{\mathrm{S}} K^+ K^-$,
      $K^0_{\mathrm{S}} K^+ \pi^-$,
      $K^0_{\mathrm{S}} \pi^+ K^-$.
       }
  \label{fig:fit-sum-PP}
\end{figure}

\begin{figure}[!htb]
  \centering
  \includegraphics[width=0.49\linewidth]{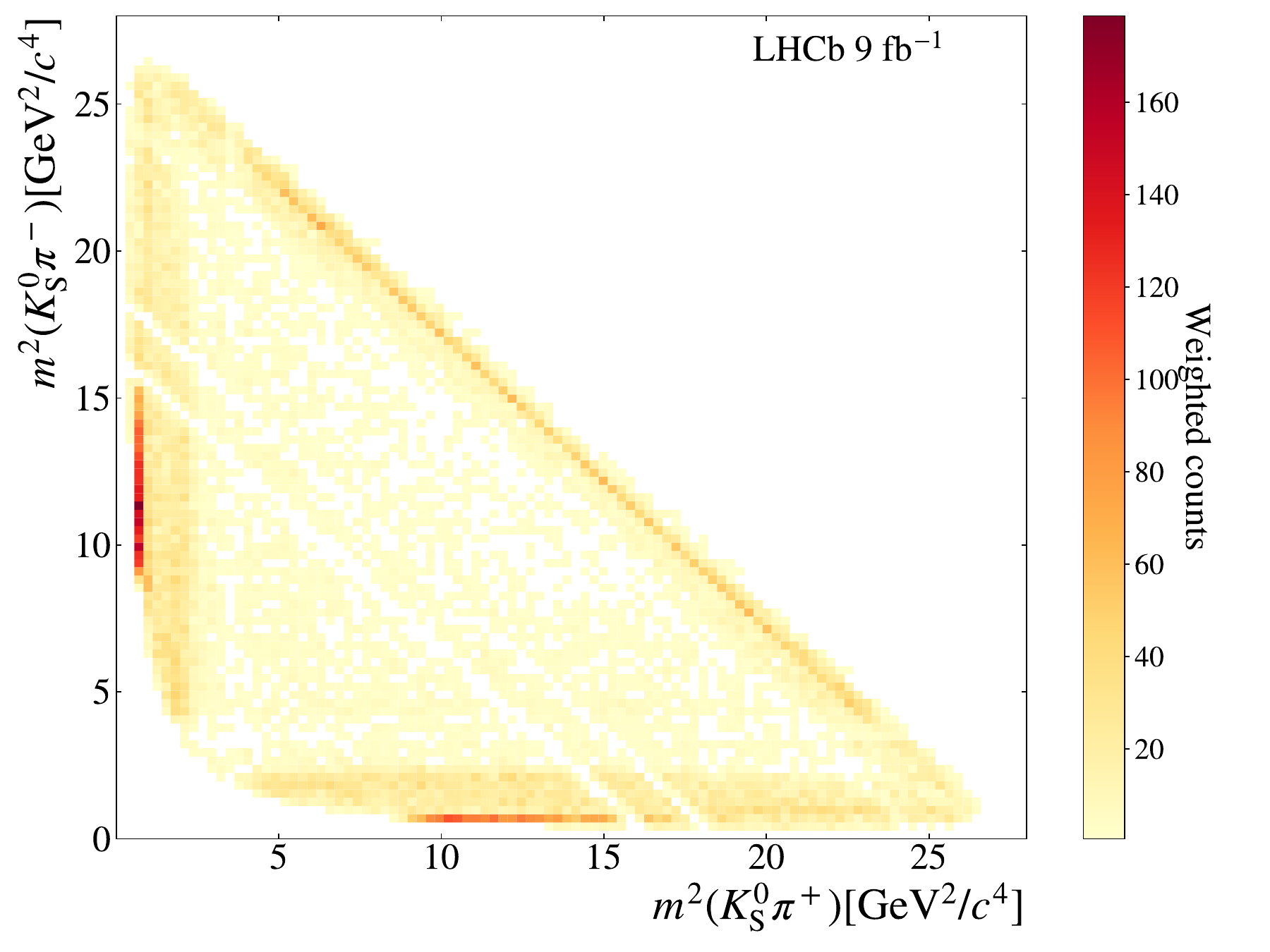}
  \includegraphics[width=0.49\linewidth]{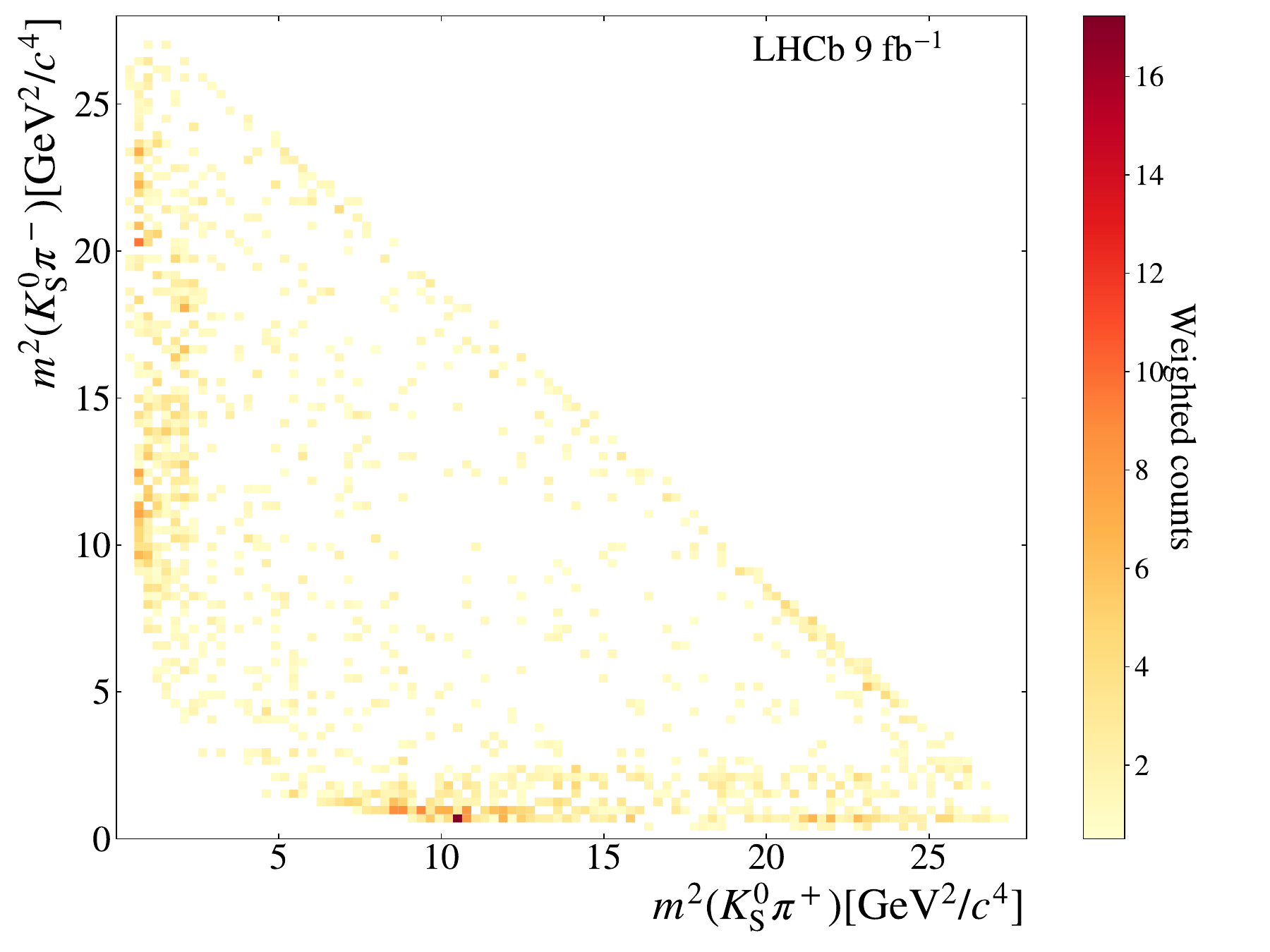}
  \includegraphics[width=0.49\linewidth]{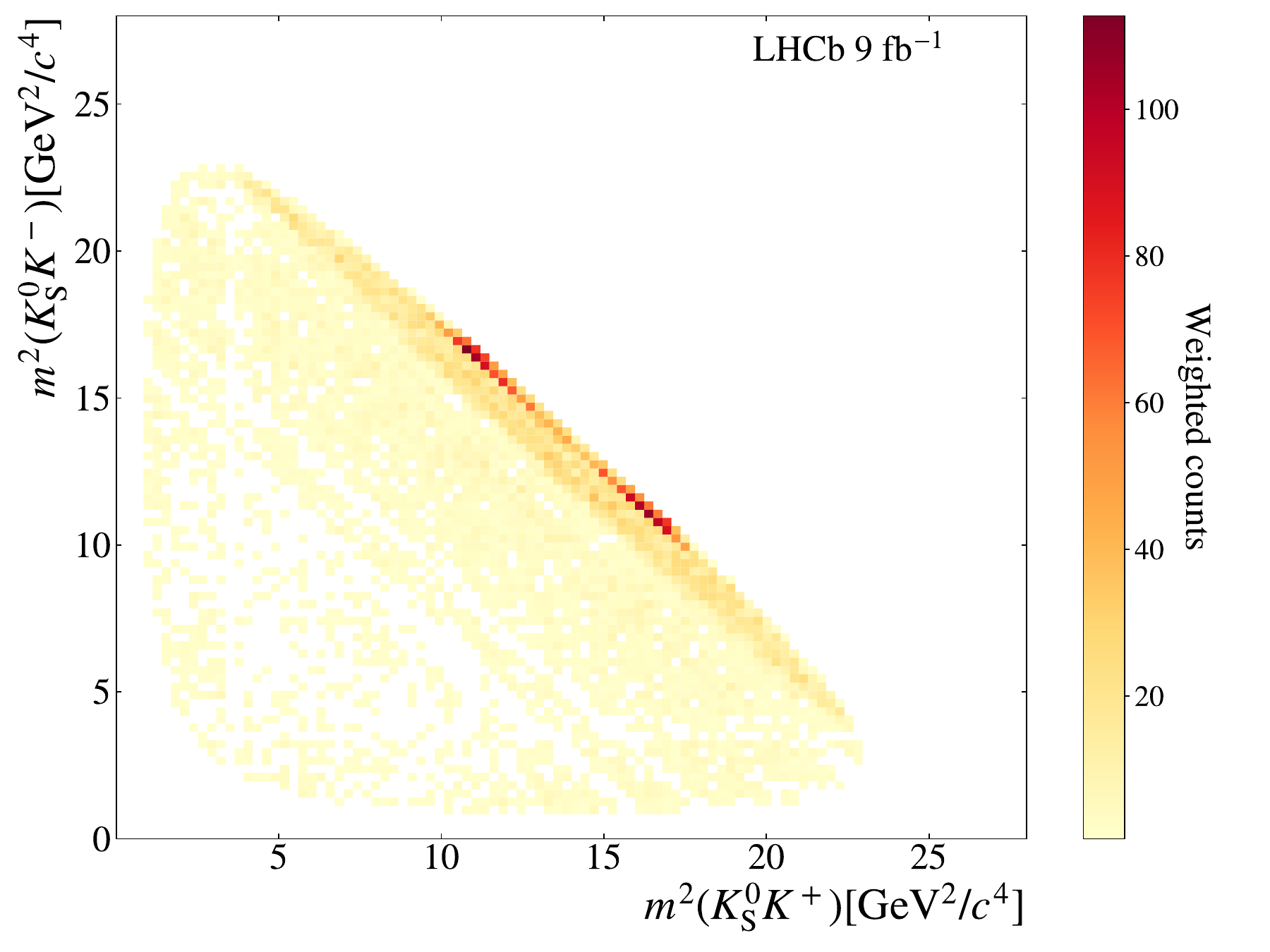}
  \includegraphics[width=0.49\linewidth]{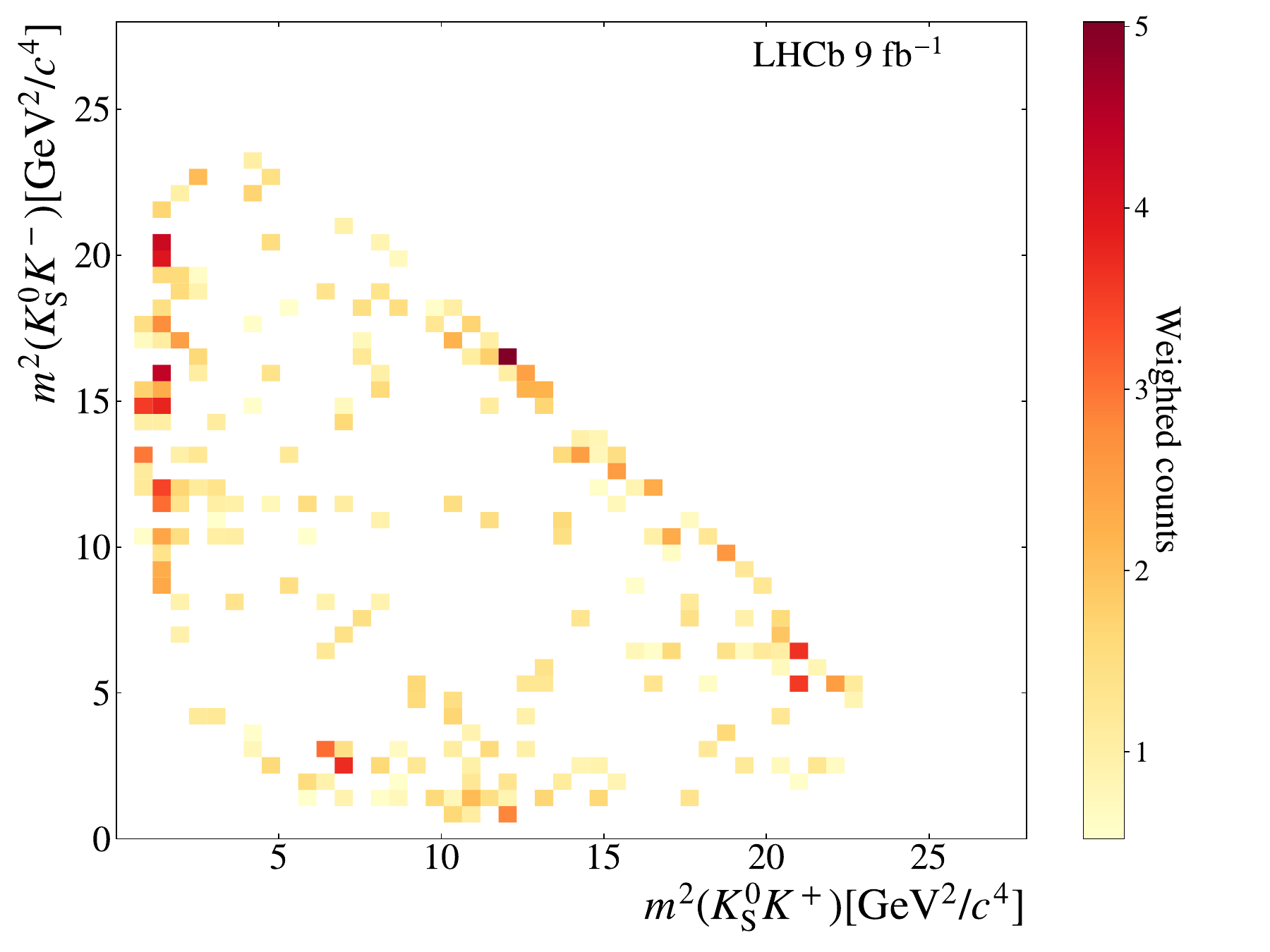}
  \includegraphics[width=0.49\linewidth]{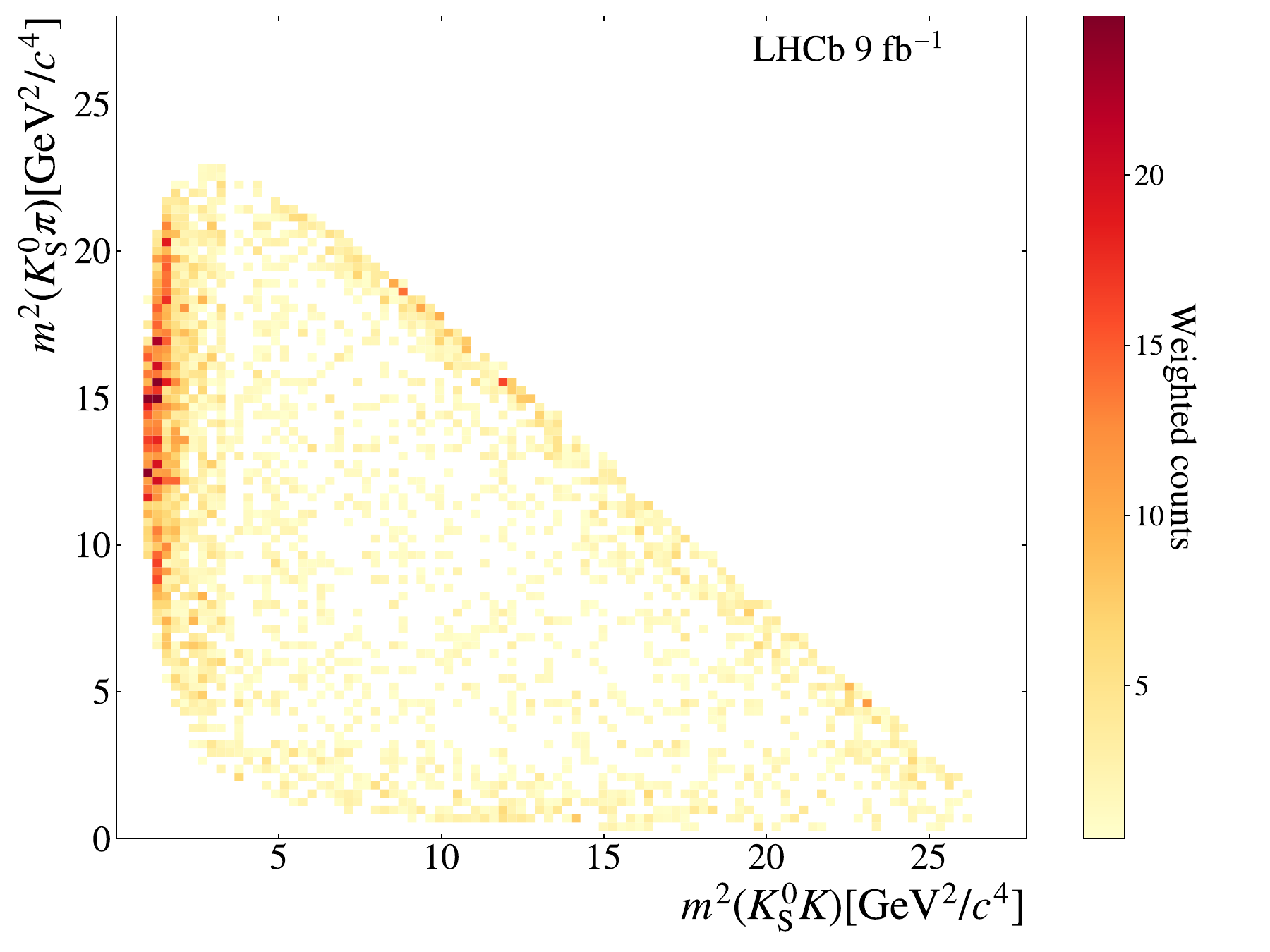}
  \includegraphics[width=0.49\linewidth]{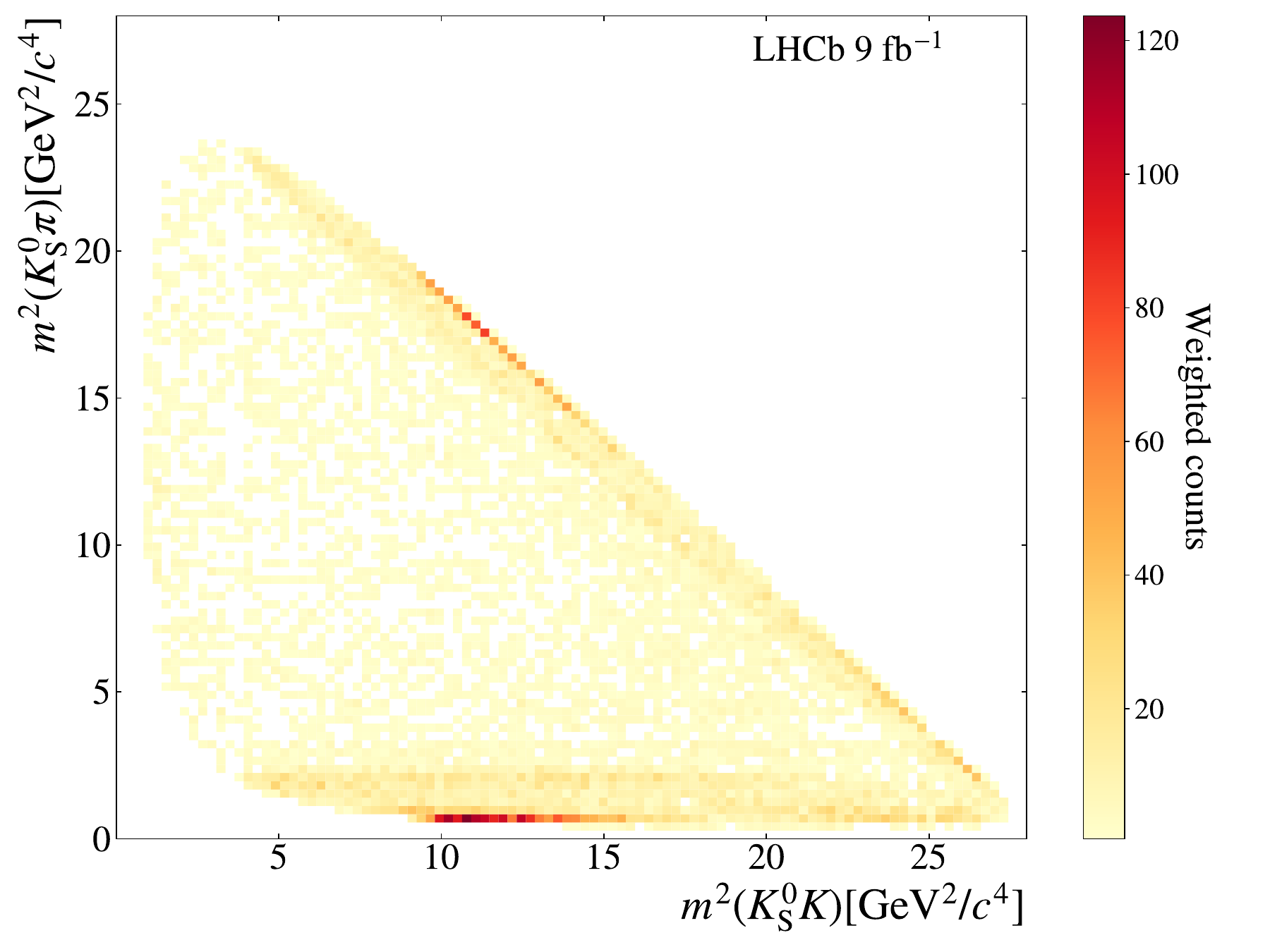}
  \caption{
    \small
      Signal Dalitz plots obtained using the signal weights computed from the results of the mass fits.
      From top to bottom, each row corresponds to (left) \Bd and (right) \Bs decays to a final state:
      $K^0_{\mathrm{S}} \pi^+ \pi^-$,
      $K^0_{\mathrm{S}} K^+ K^-$,
      $K^0_{\mathrm{S}} K^\pm \pi^\mp$.
      Bins with small negative entries are set to zero.
    }
    \label{fig:DPs}
\end{figure}

The individual fits to the subsamples can be found in Appendix~\ref{sec:app-fits-all}.
For illustration, the sum of all of the individual fits are shown, together with the combined dataset, in Fig.~\ref{fig:fit-sum-SP} for the tighter selection, and in Fig.~\ref{fig:fit-sum-PP} for the looser selection.
The total fitted signal yields, summed across all subsamples, are given in Table~\ref{tab:fit-yields-sum} with statistical uncertainties only.
In addition to the signal yields, the fit output is also used to compute signal weights
via the \sPlot\ method~\cite{Barlow:1986ek,Pivk:2004ty};
these are used for certain data-simulation corrections as described below.
They are also used to construct the signal Dalitz plots for each decay that are shown in Fig.~\ref{fig:DPs}.
The total yield of \decay{\Bs}{\KS \Kp \Km}, combining the \LL\ and \DD\ \KS reconstruction types,
is found to be $207 \pm 21$ with the tighter selection, for which the corresponding significance, based on the ratio of yield to statistical uncertainty, is approximately 10~standard deviations.

\begin{table}[!htp]
\caption{Signal yields and the associated statistical uncertainties for each final state in data, combining results from  fits to the seven data-taking periods, obtained at the two different selection working points (looser, tighter). Yields are quoted separately for the two $K^0_{\mathrm{S}}$ reconstruction types (LL, DD).}
\begin{center}
\begin{tabular}{l|cc}
Decay & Yield (looser) & Yield (tighter) \\
\midrule
$\Bd\to\KS \mathrm{(LL)}\pi^+\pi^- $ & 10930 $\pm$ 140                     & \phantom{0}6920 $\pm$ \phantom{0}90 \\
$\Bd\to\KS \mathrm{(DD)}\pi^+\pi^- $ & 25040 $\pm$ 230                     & 12360 $\pm$ 120                     \\
$\Bd\to\KS \mathrm{(LL)} K^+ K^- $   & \phantom{0}4280 $\pm$ \phantom{0}70 &                                     \\
$\Bd\to\KS \mathrm{(DD)} K^+ K^- $   & \phantom{0}9640 $\pm$ 120           &                                     \\
$\Bd\to\KS \mathrm{(LL)} \pi^+ K^- $ &                                     & \phantom{00}700 $\pm$ \phantom{0}40 \\
$\Bd\to\KS \mathrm{(DD)} \pi^+ K^- $ &                                     & \phantom{0}1430 $\pm$ \phantom{0}60 \\
$\Bd\to\KS \mathrm{(LL)} K^+ \pi^- $ &                                     & \phantom{00}576 $\pm$ \phantom{0}35 \\
$\Bd\to\KS \mathrm{(DD)} K^+ \pi^- $ &                                     & \phantom{0}1480 $\pm$ \phantom{0}60 \\
\midrule
$\Bs\to\KS \mathrm{(LL)}\pi^+\pi^- $ &                                     & \phantom{00}485 $\pm$ \phantom{0}34 \\
$\Bs\to\KS \mathrm{(DD)}\pi^+\pi^- $ &                                     & \phantom{00}980 $\pm$ \phantom{0}50 \\
$\Bs\to\KS \mathrm{(LL)} K^+ K^- $   &                                     & \phantom{000}60 $\pm$ \phantom{0}12  \\
$\Bs\to\KS \mathrm{(DD)} K^+ K^- $   &                                     & \phantom{00}147 $\pm$ \phantom{0}17  \\
$\Bs\to\KS \mathrm{(LL)} \pi^+ K^- $ & \phantom{0}2750 $\pm$ \phantom{0}70 &                                     \\
$\Bs\to\KS \mathrm{(DD)} \pi^+ K^- $ & \phantom{0}6300 $\pm$ 110           &                                     \\
$\Bs\to\KS \mathrm{(LL)} K^+ \pi^- $ & \phantom{0}2550 $\pm$ \phantom{0}70 &                                     \\
$\Bs\to\KS \mathrm{(DD)} K^+ \pi^- $ & \phantom{0}6080 $\pm$ 110           &                                     
\end{tabular}
\end{center}
\label{tab:fit-yields-sum}
\end{table}

\section{Determination of the efficiencies}
\label{sec:Efficiencies}

The following expression is used to determine branching fractions of the \BdstoKshhp decays relative to the normalisation mode \BdtoKsPiPi,
\begin{eqnarray}
\frac{\BF(\BdstoKshhp)}{\BF(\BdtoKsPiPi)} & = &
\frac{\eps^{\rm sel}_{\BdtoKsPiPi}}{\eps^{\rm sel}_{\BdstoKshhp}}
\frac{N_{\BdstoKshhp} }{ N_{\BdtoKsPiPi}} \frac{f_d}{f_{d(s)}}\,,
\label{eqn:BF-definition}
\end{eqnarray}
where $\eps^{\rm sel}_X$ denotes the average efficiency of the decay mode $X$, $N_X$ is the fitted signal yield, and $f_d$ and $f_s$ are the hadronisation fractions of a \bquark quark into a \Bz or \Bs meson, respectively.
The efficiency term includes the geometrical acceptance and the trigger, reconstruction, and selection requirements.
The ratio \fsfdinline has been measured previously at the \lhcb experiment using hadronic and semileptonic decays at different centre-of-mass energies~\cite{LHCb-PAPER-2020-046}.
When calculating the branching fraction for a given data subsample, the value of \fsfdinline used is that corresponding to the centre-of-mass energy at which the subsample was recorded.
The uncertainty on \fsfdinline is treated as being fully correlated across all subsamples.

Three-body hadronic decays such as \BdstoKshhp may proceed through quasi-two-body intermediate resonances, and may also include broad amplitudes, referred to as nonresonant, that may overlap and interfere with one another. The resulting distribution of events within the phase space is characterised as a Dalitz plot~\cite{Dalitz:1953cp}. In the case of charmless three-body $b$-meson decays, the distribution is far from uniform, typically with much higher probability density near the boundaries of the Dalitz plot (as illustrated for the decay \BstoKsKPi in Ref.~\cite{LHCB-PAPER-2018-045}). Because the trigger, reconstruction, and selection efficiencies can vary considerably across the Dalitz plot, the efficiency $\eps^{\rm sel}_X$ is obtained by a weighted average across the phase space, with weights computed from the signal distribution.
The signal distributions are determined from the data sample analysed: although for certain decay modes there are prior measurements of the Dalitz-plot distribution, these are drawn from smaller signal yields.

To simplify the description of the phase space, a transformation into a square Dalitz plot~\cite{Aubert:2005sk} is used; this has the benefit of spreading out the regions close to the kinematic boundaries where resonances are concentrated and where the efficiency varies rapidly. The square Dalitz plot is divided into bins in a $10\times10$ grid. For a data-taking period $k$, the efficiency within bin $j$ is denoted $\varepsilon_{j,k}$ and is computed from simulated events, corrected for known data-simulation differences as described below.

For each subsample $k$ of data, the fits to the $\KS h^+ h^{\prime -}$ mass spectra described in Sec.~\ref{sec:FitModel} are used to generate weights, $W_{j,k}$, using the \sPlot\ technique~\cite{Pivk:2004ty}, taking the $\KS h^+ h^{\prime -}$ mass as the discriminating variable; these weights are then used to isolate the \Bz and \Bs signals in data and determine their distributions in the square Dalitz plots.
Correlations between the Dalitz-plot variables and the discriminating variable are found to be sufficiently small to be neglected.
While this can be done for all subsamples, the distributions are best determined with larger sample sizes.
Therefore, for the purposes of averaging the efficiency, only the last three years $k^{\prime} = \{2016, 2017, 2018\}$ are used (with distributions quantified as the sum of the weights $W_{j,k'}$ in each square Dalitz bin $j$), with a given subsample only included if the statistical significance of the signal yield exceeds 2 standard deviations.
The only case where this requirement results in a subsample being excluded is in the determination of the \BstoKsKK distribution, where the 2016 LL subsample has an insignificant signal yield ($1.5 \pm 2.6$).
Each of these distributions is then corrected for efficiency in bins of the square Dalitz plot; the resulting efficiency-corrected, background-subtracted distribution may then be used to compute the average efficiency for any subsample, including those with smaller yields.
The estimate of the weighted-average efficiency for subsample $k$, determined using the distribution of subsample $k^\prime$, is given by
\begin{equation}
\bar{\varepsilon}_{k,k'} = \dfrac{\sum_j W_{j,k'} \, \varepsilon^{-1}_{j,k'} \, \varepsilon_{j,k}}{\sum_j W_{j,k'} \, \varepsilon^{-1}_{j,k'}}.
\label{ave_eps4}
\end{equation}
Note that for each data-taking period $k$, there are three independent estimates of the average efficiency, one for each of the reference data-taking periods.
These three estimates are then combined, taking the arithmetic mean without weighting (since the three reference data samples are approximately equal in size), to give single efficiency $\bar{\varepsilon}_{k}$ for each dataset $k$.

Known differences between data and simulation are corrected with calibration samples.
For the \hadp and \hadprimm hadrons, differences in tracking efficiency are corrected as a function of the track momentum and pseudorapidity~\cite{LHCb-DP-2013-002}.
An analogous correction is applied to the \KS tracking and vertex-reconstruction efficiency.
Differences in hardware-trigger response between data and simulation are determined from samples of $\Dstarp \to \Dz (\to \Km\pip) \pi^+$ decays triggered by other activity in the event, and are determined as a function of the hadron charge, \pt, calorimeter region, and the overlap between the calorimeter clusters of the final-state particles in signal decays, including a correction for the average underlying energy in the calorimeter.
Differences in particle-identification efficiency and misidentification rates between kaons and pions are determined from the same \Dstarp calibration samples; as noted in Sec.~\ref{sec:Selection}, these corrections are applied before simulated decays are processed by the PID BDT.
The tracking-efficiency corrections are of $\order(1\%)$, while those related to the trigger efficiency can be up to $10\%$.

The selection efficiency depends upon the $b$-hadron decay-time distribution.
In the case of \Bs decays, the lifetime difference between the two \CP eigenstates is significant, and at present the \CP content of the three \Bs decays is unknown.
The \Bs decay model used in the simulation assumes an effective lifetime of $1/\Gs$, where $\Gs$ is the average width of the two \CP-eigenstates of the \Bs meson.
The relative change in the \CP-averaged \Bs efficiency corresponding to a variation of the effective \Bs meson decay width of $\pm\DGs/2$ is estimated to be $\mp4\%$.

\section{Systematic uncertainties}
\label{sec:Systematics}

The branching fractions are measured relative to that of the decay \BdtoKsPiPi, as shown in
Eq.~\ref{eqn:BF-definition}. Similar selections and fit procedures
are used
for the numerator and denominator,
and
as a result, many systematic effects cancel at first order in the ratio.
The remaining uncertainties are discussed below,
grouped according to whether they influence the yield measurements ($N$) or the efficiency ($\varepsilon$);
the uncertainty on $f_s/f_d$ is treated separately.
In most cases, systematic uncertainties are first evaluated on the individual measurements
(split by data-taking period and \KS reconstruction type);
the procedure by which these individual measurements are combined, and the
systematic uncertainties on this combination,
are discussed in Sec.~\ref{sec:Sys:Combination}.

\subsection{Fit model and yield measurement uncertainties}
\label{sec:Systematics-fit}

Several assumptions are made during the fit procedure, principally the models chosen to describe different components and the choices made when constraining or fixing fit parameters. For each of the 13 assumptions discussed below, the associated systematic uncertainty is evaluated by making a different assumption with respect to the baseline model and rerunning the full analysis up to the determination of the ratios of branching fractions~(Eq.~\ref{eqn:BF-definition}). The uncertainty on a given ratio for a given subsample is taken to be the absolute value of the difference between the result obtained with the baseline method and with the modified method. This is done separately for each ratio of branching fractions (with $B^0_{(s)} \to \KS K^{\pm} \pi^{\mp}$ split by charge), each data-taking period, each \KS reconstruction type, and for the looser and tighter selections (\ie for $7 \times 7 \times 2 \times 2$ configurations).

For the signal component, systematic uncertainties are assigned based on the two following  modifications:
  replacing the signal model with a double-sided Hypatia function~\cite{MartinezSantos:2013ltf}, or
  allowing the mass difference $\Delta m$ between the \Bz and \Bs peaks to vary in the fit.

For the cross-feed component, the two following modifications are considered:
  the constraint on the ratio of yields of the two cross-feed peaks in each spectrum is removed, allowing them to vary independently; or
  the yields of cross-feed peaks are allowed to vary instead of being fully constrained when using the tighter selection.

For the partially reconstructed component, fits to representative simulated samples are used to determine the lineshape.
  In the baseline fit, the charmless decay \decay{\Bz}{K^\ast(892)^{0} \rho(770)^0} is used for \Bz and, with an offset, for \Bs decays.
  In the modified versions, a lineshape derived from fits to a sample of the hadronic decay \decay{B^+}{\Dz \pip}, with \decay{\Dz}{\KS \pip \pim}, is considered,
  first in place of the charmless decay, or second in addition to it.
  In a third variation, additional components are added to the $\KS \pip \pim$ spectrum to describe radiative decays.

For the combinatorial component, four variations are considered:
  the slope parameters for the final states $\KS \Kp \pim$ and $\KS \pip \Km$ are allowed to differ,
  the slope parameter is allowed to take positive values,
  the slope parameter is fixed entirely according to a prior fit to the upper mass sideband 5500--5750\mevcc, or
  the combinatorial model is replaced with an exponential function.

Finally, in the baseline fit procedure, very small components are suppressed because yields close to the lower bound of zero can cause fit instability: when the fit finds a component yield to be less than one event for a subsample, this is taken to be negligible and fixed to zero, and the data are refitted without that component. This procedure is instead modified in two ways: by fixing the yield to the fitted value, or by leaving the yield free.

For each subsample and configuration, the procedure above gives a set of 13 fit systematic uncertainties on each ratio of branching fractions. These 13 uncertainties are then combined in quadrature to give an uncertainty on each ratio of branching fractions for the subsample and configuration. These fit uncertainties on the individual subsamples are subsequently propagated, assuming no correlation among the subsamples, to the averaged ratios of branching fractions (see Sec.~\ref{sec:Sys:Combination}). Typically, the largest fit systematic uncertainties arise from fixing the combinatorial slope parameter from a fit to the upper mass sideband, and from using charmed hadronic decays to model the partially reconstructed component.

The systematic uncertainty due to the fit method on the combined yield of \decay{\Bs}{\KS \Kp \Km} with the tighter selection is 12.
Adding this in quadrature with the statistical uncertainty gives a yield of $207 \pm 24$, for which the corresponding significance, based on the ratio of the yield to the total uncertainty, is $8.6$~standard deviations.

\subsection{Efficiency uncertainties}

Systematic uncertainties on the efficiency ratios arise from the finite size of the samples of simulated signal events, the accuracy of the corrections for data-simulation differences, and the method used to average the efficiencies.

Uncertainties on the tracking, hardware trigger, and PID corrections arise from the finite size of the data calibration samples. For the hardware-trigger correction, an additional uncertainty arises from the sample with which the average underlying energy is determined; this uncertainty is estimated by remeasuring the average underlying energy with another data calibration sample, using identified charm events instead of identified beauty events. For the PID correction, additional uncertainties arise from the finite size of the simulation sample (used to determine the signal-track kinematic distribution for the correction), and from the nonparametric kernel-density method used to describe the PID response; the latter uncertainty is estimated by varying the kernel width.

In addition, the kinematic distributions of \Bz and \Bs mesons may differ between simulation and data; the effect of this is tested by weighting the simulated data to match the kinematic distributions in higher-yield data samples of \decay{B_{(s)}^0}{\KS h^+ h^{\prime -}} decays and recomputing the efficiency. The change in mean efficiency after reweighting is determined for 18 separate subsamples, using combinations of three decay channels (\decay{\Bd}{\KS \pip \pim}, \decay{\Bd}{\KS \Kp \Km}, \decay{\Bs}{\KS K^\pm \pi^\mp}), three data-taking periods (2016, 2017, 2018), and the two \KS reconstruction types. The effect is found to be small, and a systematic uncertainty is assessed based on the sum in quadrature of the mean and RMS of the 18 subsamples. This uncertainty is taken to be correlated between data-taking periods.

The uncertainty on the method used to compute the weighted-average efficiency $\bar{\varepsilon}_{k}$ is based on the spread among the three independent estimates 
$\bar{\varepsilon}_{k,k'}$
(see Eq.~\ref{ave_eps4}), quantified as the RMS of
$\{ \bar{\varepsilon}_{k,k'} - \bar{\varepsilon}_{k} \}$
among all combinations $k,k'$ for a given decay mode and \KS reconstruction type.
(Note that there are $7 \times 3$ such combinations, for the seven data-taking periods $k$ and the three reference periods $k'$.) While this does mean that some subsamples are reused in the uncertainty estimate it has the important advantage that the systematic uncertainty is stable across data-taking periods and is only mildly affected by statistical fluctuations in the smaller subsamples. This uncertainty is  correlated between data-taking periods.

The average efficiency is also influenced by the choice of binning scheme for the square Dalitz plot. Instead of the baseline $10 \times 10$ binning, six other $n \times n$ grids are used (for $n = 7,8,9,11,12,13$) and, for each grid, the average efficiency is recomputed and the difference taken with respect to that of the baseline $n=10$ grid. This is done for each of the seven data-taking periods, for a total of 42 values, and the RMS of the distribution is taken as an estimate of the associated systematic uncertainty. This generates one uncertainty per decay mode, \KS reconstruction type, and selection working point (looser/tighter), and is thus correlated between data-taking periods.

\subsection{Combination}
\label{sec:Sys:Combination}

For each of the five ratios of branching fractions $R_i$, fourteen separate measurements $R_{ij}$ are made
(from the seven data-taking periods and two \KS reconstruction types).
In general, the agreement between all of the measurements is excellent.
Only a single measurement of one branching fraction ratio shows a significant disagreement with respect to the other thirteen measurements.
It is therefore assumed to be due to a fluctuation.
The fourteen measurements of each branching fraction ratio are combined in a weighted average:
\begin{equation}
    \overline{R}_i =  \left[ \sum_{j} w_{ij} \right]^{-1}  \sum_{j} R_{ij} \, w_{ij}
    .
    \label{eq:sys-weighted-average}
\end{equation}
The weight used for each measurement is $w_{ij} = (\sigma_{ij})^{-2}$, where $\sigma_{ij}$ is
the sum in quadrature of the statistical and uncorrelated systematic uncertainties on the ratio $R_{ij}$.

The uncertainties on the individual measurements $R_{ij}$ are propagated to the combinations $\overline{R}_i$. Other than as noted above, systematic uncertainties are assumed to be uncorrelated between subsamples and are added in quadrature.
The uncertainties are summarised in Table~\ref{tab:sys-summary}.
The largest sources of systematic uncertainty are the modelling of the fit components and the method to account for the efficiency variation over the Dalitz plot.

\begin{table}[t]
\caption{
  Summary of the relative systematic uncertainties on the ratios of branching fractions after taking the weighted average of the subsamples, as described in Sec.~\ref{sec:Sys:Combination}.
  For each column, the decay mode in the numerator of the ratio of branching fractions is indicated by the initial $b$~hadron and the $h^+ h^{\prime-}$ combination.
  An asterisk indicates that the class includes uncertainties treated as correlated between different data-taking periods.
}
\label{tab:sys-summary}
\begin{center}
\begin{tabular}{l|rrrrr}
Source [\textperthousand] & $\Bd, \Kp\Km$ & $\Bd, K^\pm \pi^\mp$ & $\Bs, \pip\pim$ & $\Bs, \Kp\Km$ & $\Bs, K^\pm \pi^\mp$  \\
\midrule 
Fit model         & 7.8    & 28.6   & 40.4   & 58.8  & 5.8    \\ 
Tracking          & 0.03  & $0.04$ & $<0.01$ & 0.07   & $0.03$ \\ 
Hardware trigger  & 1.3    & 1.0    & 1.0    & 1.4   & 0.7    \\ 
PID               & 2.3    & 5.4    & $<0.1$ & 6.2   & 1.8    \\ 
Kin. reweighting* & 6.2    & 4.4    & 6.0    & 6.3   & 4.4    \\ 
Dalitz plot*      & 27.3   & 22.6   & 37.3   & 57.2 & 15.3   \\ 
\midrule 
Total systematic  & 29     & 37     & 55     & 82   & 17     \\ 
Statistical       & 12     & 26     & 42     & 135   & 11     \\ 
$f_s/f_d$*        & ---    & ---    & 31     & 31    & 31     \\ 
\end{tabular}
\end{center}
\end{table}

\section{Results and conclusion}
\label{sec:Results}

A dataset recorded by the LHCb detector between 2011 and 2018, corresponding to an integrated luminosity of 9\invfb, is used to study the \BdstoKshhp decay modes.
The branching fraction of each decay is measured relative to that of \BdtoKsPiPi.
The ratios of branching fractions are determined independently for the two \KS reconstruction categories and seven data-taking periods, with all such determinations being in agreement.
They are then combined in a weighted average, and 
the results obtained are
\begin{alignat*}{6} 
&\frac{\mathcal{B}(\Bd\to\KS K^+ K^-)}{\mathcal{B}(\Bd\to\KS\pi^+\pi^-)} &&= 0.578 &&\pm 0.007 \, \mathrm{(stat)} &&\pm 0.017 \, \mathrm{(syst)}\,, &&\\ 
&\frac{\mathcal{B}(\Bd\to\KS K^\pm\pi^\mp)}{\mathcal{B}(\Bd\to\KS\pi^+\pi^-)} &&= 0.1363 &&\pm 0.0035 \, \mathrm{(stat)} &&\pm 0.0051 \, \mathrm{(syst)}\,, &&\\ 
&\frac{\mathcal{B}(\Bs\to\KS\pi^+\pi^-)}{\mathcal{B}(\Bd\to\KS\pi^+\pi^-)} &&= 0.269 &&\pm 0.011 \, \mathrm{(stat)} &&\pm 0.015 \, \mathrm{(syst)} && \pm 0.008 \, \mathrm{(} f_s/f_d \mathrm{)}\,,\\ 
&\frac{\mathcal{B}(\Bs\to\KS K^+ K^-)}{\mathcal{B}(\Bd\to\KS\pi^+\pi^-)} &&= 0.0303 &&\pm 0.0041 \, \mathrm{(stat)} &&\pm 0.0025 \, \mathrm{(syst)} && \pm 0.0009 \, \mathrm{(} f_s/f_d \mathrm{)}\,,\\ 
&\frac{\mathcal{B}(\Bs\to\KS K^\pm\pi^\mp)}{\mathcal{B}(\Bd\to\KS\pi^+\pi^-)} &&= 1.818 &&\pm 0.021 \, \mathrm{(stat)} &&\pm 0.031 \, \mathrm{(syst)} && \pm 0.056 \, \mathrm{(} f_s/f_d \mathrm{)}\,.
   \end{alignat*}
The  decay \BstoKsKK is observed for the first time, with a significance exceeding six standard deviations.
All measurements of branching fractions are in good agreement with the earlier \lhcb determinations~\cite{LHCb-PAPER-2013-010}, which they supersede.
The measurement of the relative branching fractions of \BdtoKsKPi and \BdtoKsKK are consistent with the world-average results~\cite{HFLAV23,PDG2024}.
At present the \CP content of the three \Bs decays is unknown but once it is established, \eg by a future amplitude analysis, the branching fraction ratios can be corrected for the modified efficiency within the estimated maximal deviation of $\pm4\%$, corresponding to a variation of the effective \Bs meson decay width of $\mp\DGs/2$.

Using the external known value (excluding input from previous LHCb measurements that used a subset of the data considered here), ${\cal B}(\BdtoKzPiPi) = (4.96 \pm 0.20) \times 10^{-5}$~\cite{PDG2024}, the measured time-integrated branching fractions are
\begin{alignat*}{6} 
    &\mathcal{B}(\Bd\to\Kz K^+ K^-)      &&= [28.6            &&\pm 0.4 \,            &&\mathrm{(stat)} \pm 0.8 \,            &&\mathrm{(syst)} \pm 1.2\,            &&(\mathrm{ext}) ] \times 10^{-6},\\ 
    &\mathcal{B}(\Bd\to\!\KorKbarz K^\pm\pi^\mp) &&= [\phantom{0}6.76 &&\pm 0.17 \,           &&\mathrm{(stat)} \pm 0.25 \,           &&\mathrm{(syst)} \pm 0.27\,           &&(\mathrm{ext}) ] \times 10^{-6},\\ 
    &\mathcal{B}(\Bs\to\Kzb \pi^+\pi^-)    &&= [13.3\phantom{0} &&\pm 0.6\phantom{0} \, &&\mathrm{(stat)} \pm 0.7\phantom{0} \, &&\mathrm{(syst)} \pm 0.7\phantom{0}\, &&(\mathrm{ext}) ] \times 10^{-6},\\ 
    &\mathcal{B}(\Bs\to\Kzb K^+ K^-)      &&= [\phantom{0}1.50 &&\pm 0.20 \,           &&\mathrm{(stat)} \pm 0.12 \,           &&\mathrm{(syst)} \pm 0.08\,           &&(\mathrm{ext}) ] \times 10^{-6},\\ 
    &\mathcal{B}(\Bs\to\!\KorKbarz K^\pm\pi^\mp) &&= [90.2\phantom{0} &&\pm 1.0\phantom{0} \, &&\mathrm{(stat)} \pm 1.5\phantom{0} \, &&\mathrm{(syst)} \pm 4.6\phantom{0}\, &&(\mathrm{ext}) ] \times 10^{-6},
\end{alignat*}
where `ext' denotes the combined uncertainties on $f_s/f_d$ and ${\cal B}(\BdtoKzPiPi)$.
These results agree with the available predictions for these channels~\cite{Cheng:2013dua,Cheng:2014uga,Li:2014fla,Li:2014oca,Wang:2016rlo,Li:2017mao}, and with previous experimental determinations.

\section*{Acknowledgements}
%
%
\noindent We express our gratitude to our colleagues in the CERN
accelerator departments for the excellent performance of the LHC. We
thank the technical and administrative staff at the LHCb
institutes.
We acknowledge support from CERN and from the national agencies:
ARC (Australia);
CAPES, CNPq, FAPERJ and FINEP (Brazil); 
MOST and NSFC (China); 
CNRS/IN2P3 and CEA (France);  
BMFTR, DFG and MPG (Germany);
INFN (Italy); 
NWO (Netherlands); 
MNiSW and NCN (Poland); 
MCID/IFA (Romania); 
MICIU and AEI (Spain);
SNSF and SER (Switzerland); 
NASU (Ukraine); 
STFC (United Kingdom); 
DOE NP and NSF (USA).
We acknowledge the computing resources that are provided by ARDC (Australia), 
CBPF (Brazil),
CERN, 
IHEP and LZU (China),
IN2P3 (France), 
KIT and DESY (Germany), 
INFN (Italy), 
SURF (Netherlands),
Polish WLCG (Poland),
IFIN-HH (Romania), 
PIC (Spain), CSCS (Switzerland), 
GridPP (United Kingdom),
and NSF (USA).  
We are indebted to the communities behind the multiple open-source
software packages on which we depend.
Individual groups or members have received support from
Key Research Program of Frontier Sciences of CAS, CAS PIFI, CAS CCEPP (China); 
Minciencias (Colombia);
EPLANET, Marie Sk\l{}odowska-Curie Actions, ERC and NextGenerationEU (European Union);
A*MIDEX, ANR, IPhU and Labex P2IO, and R\'{e}gion Auvergne-Rh\^{o}ne-Alpes (France);
Alexander-von-Humboldt Foundation (Germany);
ICSC (Italy); 
Severo Ochoa and Mar\'ia de Maeztu Units of Excellence, GVA, XuntaGal, GENCAT, InTalent-Inditex and Prog.~Atracci\'on Talento CM (Spain);
the Leverhulme Trust, the Royal Society and UKRI (United Kingdom).

\clearpage

{\noindent\normalfont\bfseries\Large Appendices}

\appendix
\section{Fits to all subsamples}
\label{sec:app-fits-all}

\begin{figure}[htb]
\begin{center}
\includegraphics[width=\AppPlotWidth\textwidth]{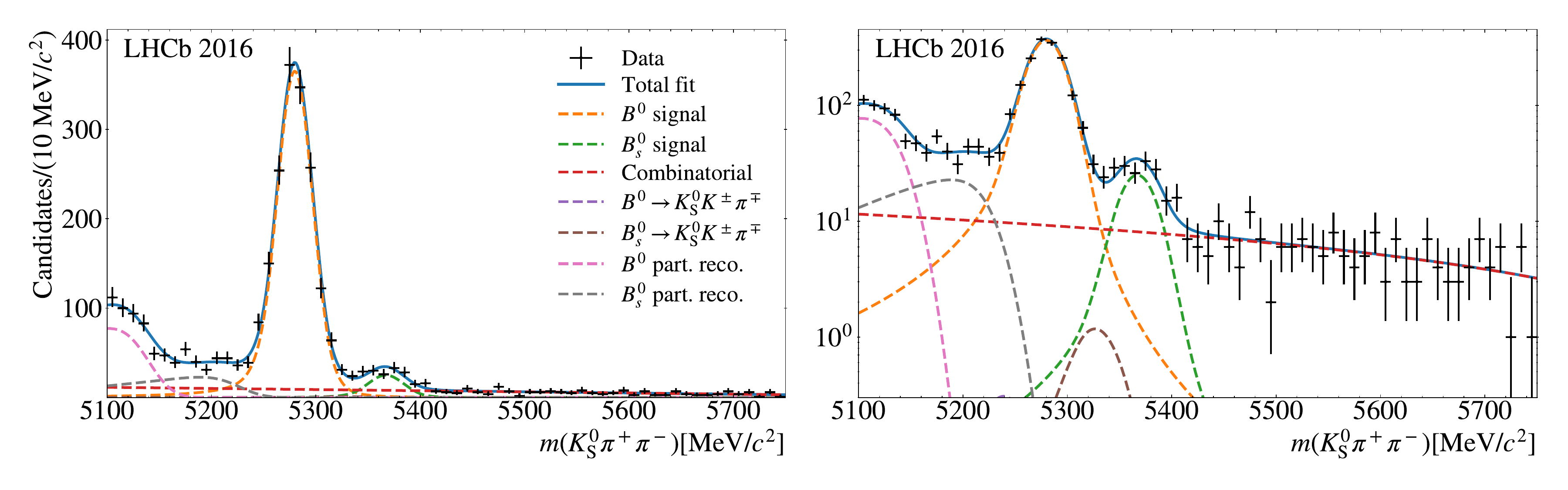}
\includegraphics[width=\AppPlotWidth\textwidth]{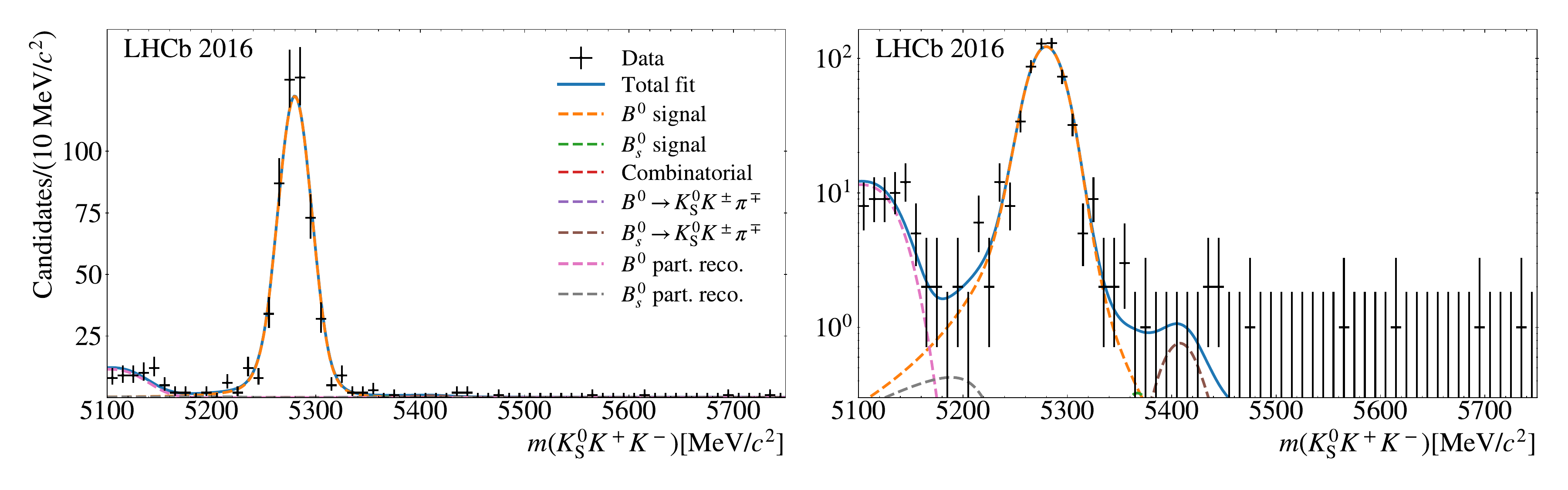}
\includegraphics[width=\AppPlotWidth\textwidth]{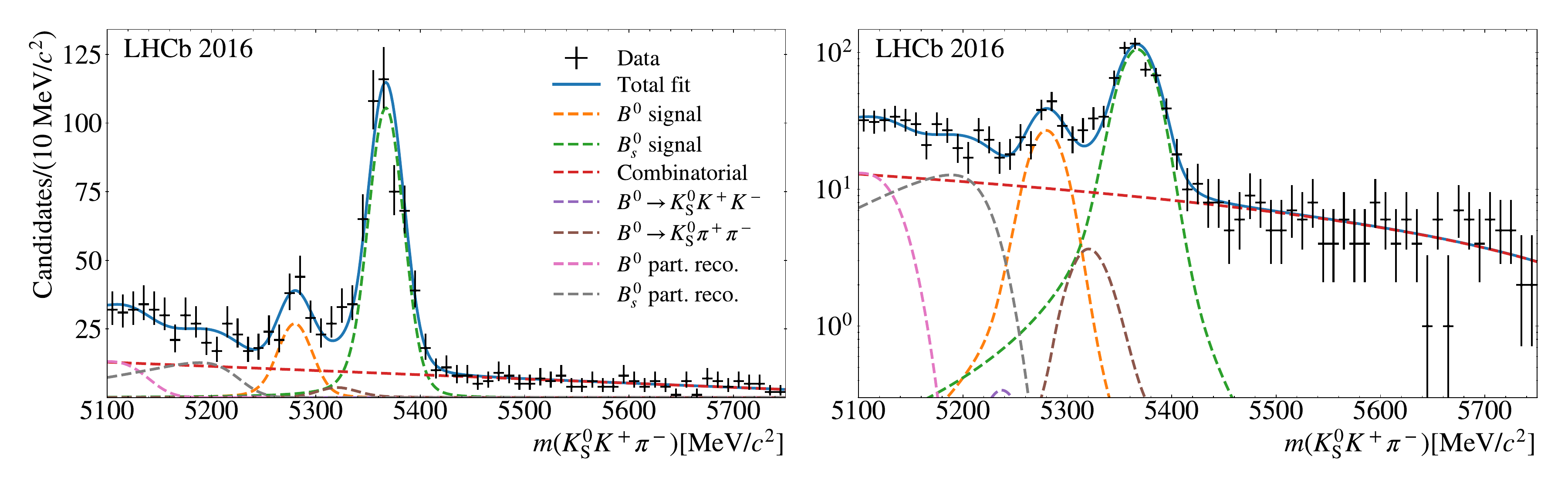}
\includegraphics[width=\AppPlotWidth\textwidth]{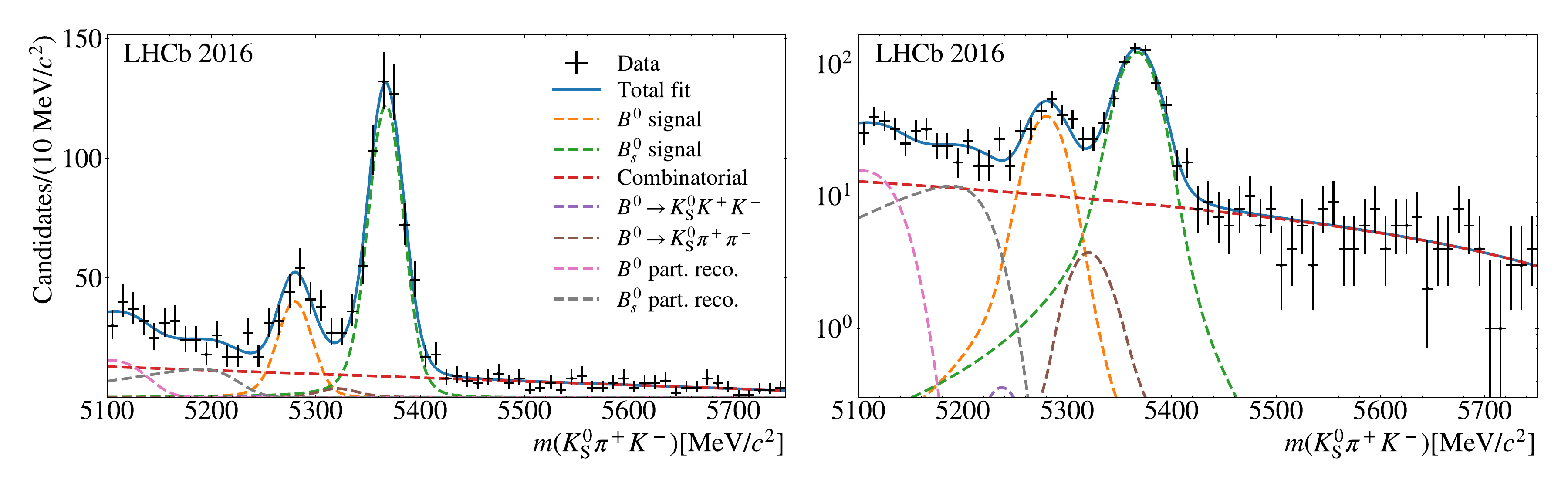}
\end{center}
\caption{\small
      Mass distributions of $K^0_{\mathrm{S}} h^+ h^{\prime -}$ candidates that pass the 
      tighter selection, for the 
      2016 running period and the 
      long $K^0_{\mathrm{S}}$ reconstruction type.
      From top to bottom, each row corresponds to a different final state:
      $K^0_{\mathrm{S}} \pi^+ \pi^-$,
      $K^0_{\mathrm{S}} K^+ K^-$,
      $K^0_{\mathrm{S}} K^+ \pi^-$,
      $K^0_{\mathrm{S}} \pi^+ K^-$.
}   
\label{fig:fit:2016:LL:SecondaryPeak}
\end{figure}

For each ratio of branching fractions, 14 separate measurements $R_{ij}$ are made and a weighted average $\overline{R}_i$ is computed, as described in Sec.~\ref{sec:Sys:Combination}. The same set of measurements may also be used to compute a $\chi^2$,
\begin{equation}
  \chi^2_i = \sum_{j} \left( 
    \frac{R_{ij} - \overline{R}_i}{\sigma_{ij}}
    \right)^2
    ,
  \label{eq:app:chisq}
\end{equation}
where $\sigma_{ij}$ is
the sum in quadrature of the statistical and uncorrelated systematic uncertainties on the ratio $R_{ij}$,
implicitly assuming Gaussian uncertainties.
The results are given in Table~\ref{tab:app:chisq}, along with the associated $p$-values. All indicate good agreement, except for the ratio $\mathcal{B}(\Bs\to\KS K^+ K^-)/\mathcal{B}(\Bd\to\KS\pi^+\pi^-)$. For the latter, the largest contribution ($9.6$) comes from the 2016 data-taking period with the \LL\ \KS reconstruction type; all other subsamples are mutually consistent (the remaining \chisqndf would be $19.84/13$, corresponding to a $p$-value of 10\%). The fit to the $\KS \Kp \Km$ spectrum for this sample is shown in Fig.~\ref{fig:fit:2016:LL:SecondaryPeak}.
In this instance, both the fitted yield of $\Bs \to \KS \Kp \Km$ ($1.5 \pm 2.6$ candidates) and the background in the peak region are small, with the consequence that the estimated statistical uncertainty is also low, enhancing the $\chi^2$ contribution. This effect is reproduced in simulated pseudoexperiments emulating the 2016 LL $\KS \Kp \Km$ mass spectrum, in which the \Bs signal yield is drawn from a Poisson distribution where the expected signal yield is determined from the branching ratio averaged across all subsamples, and expected yields of other components are taken from the fit to data. In fits to the pseudoexperiments, downward fluctuations in the fitted \Bs yield are found to be correlated with lower statistical uncertainties on the yield, and hence larger normalised residuals (pulls), with the effect becoming strongly nonlinear for fitted yields in this spectrum below five units.

\begin{table}[hbt]
  \caption{For each measured ratio of branching fractions, the estimated $\chi^2$ for compatibility among subsamples, the associated number of degrees of freedom (ndf), and the corresponding $p$-value assuming Gaussian uncertainties.}
  \label{tab:app:chisq}
  \begin{center}
    \begin{tabular}{ccc}
      Ratio & \chisqndf & $p$-value \\ \midrule \vspace{2pt}
      $\mathcal{B}(\Bd\to\KS K^+ K^-)/\mathcal{B}(\Bd\to\KS\pi^+\pi^-)$       & $17.9/13$ & 16\% \\ \vspace{2pt}
      $\mathcal{B}(\Bd\to\KS K^\pm\pi^\mp)/\mathcal{B}(\Bd\to\KS\pi^+\pi^-)$  & $\phantom{0}6.91/13$ & 91\% \\ \vspace{2pt}
      $\mathcal{B}(\Bs\to\KS\pi^+\pi^-)/\mathcal{B}(\Bd\to\KS\pi^+\pi^-)$     & $\phantom{0}10.5/13$ & 65\% \\ \vspace{2pt}
      $\mathcal{B}(\Bs\to\KS K^+ K^-)/\mathcal{B}(\Bd\to\KS\pi^+\pi^-)$       & $33.0/13$ & 0.17\% \\ 
      $\mathcal{B}(\Bs\to\KS K^\pm\pi^\mp)/\mathcal{B}(\Bd\to\KS\pi^+\pi^-)$  & $\phantom{0}9.50/13$ & 73\%
    \end{tabular}
  \end{center}
\end{table}

\clearpage

\section{Definition of the signal model}
\label{sec:app-lineshapes}

In this analysis, the probability density function (PDF) of a Crystal Ball function~\cite{Skwarnicki:1986xj}, as implemented in \roofit within \root~\cite{Brun:1997pa}, is defined as follows:
\begin{equation}
    P_{\mathrm{CB}}(m ; \mu, \sigma, n, \alpha) \equiv N \left\{
    \begin{aligned}
        e^{-t^2/2}
        &\mathrm{\;\;\;\;\;\;if\;\;} t > -|\alpha| \,, \cr
        \frac{1}{ 
          \left( 
            \frac{n}{|\alpha|} - |\alpha| - t
          \right)^{n}
        }
        \left(\frac{n}{|\alpha|}\right)^{n}
        e^{-\alpha^2/2}
        &\mathrm{\;\;\;\;\;\;if\;\;} t\leq -|\alpha| \,, \cr
    \end{aligned}\right.
\end{equation}
where
\begin{equation}
t = \left\{
  \begin{array}{cl}
  \phantom{-}(m-\mu)/\sigma & \mbox{if $\alpha>0$}\,, \\
            -(m-\mu)/\sigma & \mbox{if $\alpha<0$}\,, \\
  \end{array}\right.
\end{equation}
and where $N$ is a normalisation factor such that the PDF has unit integral within the fit range. The signal PDF is then defined as
\begin{equation}
  P_{\mathrm{sig}}(m ; \mu, \sigma, n_L, n_R, \alpha_L, \alpha_R, f_L) \equiv
         f_L  P_{\mathrm{CB}}(m ; \mu, \sigma, n_L, \alpha_L)
    + (1-f_L) P_{\mathrm{CB}}(m ; \mu, \sigma, n_R, -\alpha_R)
    .
\end{equation}

Fits to signal candidates in simulation are used to obtain a set of signal shape parameters, some of which are used in the fit to data as described below. Simultaneous fits are used to share parameters between different decay modes and different \KS reconstruction types for a given data-taking period (but not between different data-taking periods). The 22 shape parameters determined for each data-taking period are:
\begin{itemize}
  \item Two peak position parameters $\mu(\Bz)$ and $\mu(\Bs)$;
  \item One width parameter $\sigma(\Bz \to \KS \pip \pim, \mathrm{DD})$ for the decay $\Bz \to \KS \pip \pim$ with the downstream \KS reconstruction;
  \item Four width ratio parameters, $\left\{ r_{\Bs}, \, r_{K\pi}, \, r_{KK}, \, r_{\mathrm{LL}} \right\}$, which together define the width of the other signal peaks relative to $\sigma(\Bz \to \KS \pip \pim, \mathrm{DD})$ under the assumption of factorisation:
  \begin{eqnarray*}
     \sigma(\Bs \to \KS \pip \pim, \mathrm{DD}) &=& \sigma(\Bz \to \KS \pip \pim, \mathrm{DD}) \, r_{\Bs} \,, \\
     \sigma(\Bz \to \KS \pip \pim, \mathrm{LL}) &=& \sigma(\Bz \to \KS \pip \pim, \mathrm{DD}) \, r_{\mathrm{LL}} \,, \\
     \sigma(\Bs \to \KS \pip \pim, \mathrm{LL}) &=& \sigma(\Bz \to \KS \pip \pim, \mathrm{DD}) \, r_{\Bs} \, r_{\mathrm{LL}} \,,  \\
     \sigma(\Bz \to \KS K^{\pm} \pi^{\mp}, \mathrm{DD}) &=& \sigma(\Bz \to \KS \pip \pim, \mathrm{DD}) \, r_{K\pi} \,, \\
     \sigma(\Bs \to \KS K^{\pm} \pi^{\mp}, \mathrm{DD}) &=& \sigma(\Bz \to \KS \pip \pim, \mathrm{DD}) \, r_{\Bs} \, r_{K\pi} \,, \\
     \sigma(\Bz \to \KS K^{\pm} \pi^{\mp}, \mathrm{LL}) &=& \sigma(\Bz \to \KS \pip \pim, \mathrm{DD}) \, r_{K\pi} \, r_{\mathrm{LL}} \,, \\
     \sigma(\Bs \to \KS K^{\pm} \pi^{\mp}, \mathrm{LL}) &=& \sigma(\Bz \to \KS \pip \pim, \mathrm{DD}) \, r_{\Bs} \, r_{K\pi} \, r_{\mathrm{LL}} \,, \\
     \sigma(\Bz \to \KS \Kp \Km, \mathrm{DD}) &=& \sigma(\Bz \to \KS \pip \pim, \mathrm{DD}) \, r_{KK} \,, \\
     \sigma(\Bs \to \KS \Kp \Km, \mathrm{DD}) &=& \sigma(\Bz \to \KS \pip \pim, \mathrm{DD}) \, r_{\Bs} \, r_{KK} \,, \\
     \sigma(\Bz \to \KS \Kp \Km, \mathrm{LL}) &=& \sigma(\Bz \to \KS \pip \pim, \mathrm{DD}) \, r_{KK} \, r_{\mathrm{LL}} \,, \\
     \sigma(\Bs \to \KS \Kp \Km, \mathrm{LL}) &=& \sigma(\Bz \to \KS \pip \pim, \mathrm{DD}) \, r_{\Bs} \, r_{KK} \, r_{\mathrm{LL}} \,; 
  \end{eqnarray*}
  \item Six left-hand tail parameters, which are different between final states but shared between \KS reconstruction types and between the \Bz and \Bs peaks for a given final state:
    $\left\{ \alpha_L(\KS\pip\pim), \, \alpha_L(\KS K^{\pm} \pi^{\mp}), \, \alpha_L(\KS\Kp\Km) \right\}$
    and
    $\left\{ n_L(\KS\pip\pim), \, n_L(\KS K^{\pm} \pi^{\mp}), \, n_L(\KS\Kp\Km) \right\}$;
  \item Six parameters to define the right-hand tail, each expressed as a ratio with respect to the corresponding left-hand tail parameter give above,
  \eg $\alpha_R(\KS\pip\pim) / \alpha_L(\KS\pip\pim)$,
  and which are likewise different between final states but shared between \KS reconstruction types and between the \Bz and \Bs peaks for a given final state;
  \item Three parameters describing the fraction of the signal in the component with the left-hand tail, which are different between final states but shared between \KS reconstruction types and between the \Bz and \Bs peaks for a given final state:
  $\left\{ f_L(\KS\pip\pim), \, f_L(\KS K^{\pm} \pi^{\mp}), \, f_L(\KS\Kp\Km) \right\}$.
\end{itemize}

In the simultaneous fits to data, for each data-taking period most shape parameters are fixed according to the corresponding fits to simulation. Specifically, all of the width ratio parameters ($r$), all of the tail parameters ($n_L, n_R, \alpha_L, \alpha_R$), and all of the fraction parameters ($f_L$) are fixed. The width parameter $\sigma(\Bz \to \KS \pip \pim, \mathrm{DD})$ is allowed to vary, to take into account potential differences in resolution between data and simulation. Similarly, one peak position parameter $\mu(\Bz)$ is allowed to vary so as to take into account potential differences in mass scale between data and simulation; the other peak position parameter is then derived as $\mu(\Bs) = \mu(\Bz) + \Delta m$, where $\Delta m$ is fixed to the known difference in mass between the two states~\cite{PDG2024}. All 16 signal yields (for the four final states with $\KS\Kp\pim$ and $\KS\pip\Km$ separated by charge, two \KS reconstruction types, and \Bz and \Bs) for each data-taking period are free parameters.

Note that the fits described above are carried out separately at the two selection working points (loose and tight), since signal shapes and yields depend upon the selection.


\section{Correlation matrices}
\label{sec:corr-matrices}

\subsection{Statistical correlation matrix}

The statistical correlations between the ratios of branching fractions are reported in Table~\ref{tab:stat-correlations}.
Some branching fraction results are determined from the tight selection and others from the loose selection.
The correlation matrix is computed from the tight selection, since those data are a subsample of those in the loose selection.

The correlation among yields is obtained directly from the fit and is found to be low (below $1\%$ for most cases).
Then the correlation is propagated to the ratio of yields and finally to the ratio of branching fractions.
The final correlation obtained mainly results from the common denominator decay mode \BdtoKsPiPi.

\begin{table}[b]
\centering
\caption{Statistical correlations, calculated as described in the text, between the ratios of branching fractions \BF(\BdstoKshhp)/\BF(\BdtoKsPiPi), where the numerator decay mode is given by the row and column headings.}
\label{tab:stat-correlations}
\resizebox{\textwidth}{!}{
$
\bordermatrix{
    & \Bd \to \KS \Kp \Km & \Bd \to \KS K^{\pm} \pi^{\mp}&  \Bs \to \KS \pip \pim& \Bs \to \KS \Kp \Km & \Bs \to \KS K^{\pm} \pi^{\mp}   \cr
\Bd \to \KS \Kp \Km & 1.00 & 0.16 & 0.06 & 0.03 & 0.33 \cr
\Bd \to \KS K^{\pm} \pi^{\mp} & 0.16 & 1.00 & 0.04 & 0.01 & 0.29 \cr
\Bs \to \KS \pip \pim & 0.06 & 0.04 & 1.00 & 0.006 & 0.07 \cr
\Bs \to \KS \Kp \Km  & 0.03 &0.01 & 0.006 & 1.00 & 0.03 \cr
\Bs \to \KS K^{\pm} \pi^{\mp} & 0.33 & 0.29 & 0.07 & 0.03 & 1.00 \cr
}
$
}
\end{table}

The statistical covariance matrix is shown in Table~\ref{tab:stat-covariance}. It is obtained from Table~\ref{tab:stat-correlations} using the statistical uncertainties on branching fractions reported in Sec.~\ref{sec:Results}.

\begin{table}[bt]
\centering
\caption{Statistical covariance matrix between the ratios of branching fractions \BF(\BdstoKshhp)/\BF(\BdtoKsPiPi), where the numerator decay mode is given by the row and column headings.}
\label{tab:stat-covariance}
\resizebox{\textwidth}{!}{
$
\bordermatrix{
    & \Bd \to \KS \Kp \Km & \Bd \to \KS K^{\pm} \pi^{\mp}&  \Bs \to \KS \pip \pim& \Bs \to \KS \Kp \Km & \Bs \to \KS K^{\pm} \pi^{\mp}   \cr
\Bd \to \KS \Kp \Km & 5.11 & 0.39 & 0.51 & 0.09 &  4.83 \cr
\Bd \to \KS K^{\pm} \pi^{\mp} & 0.39 & 1.21 & 0.17 & 0.02 &2.10 \cr
\Bs \to \KS \pip \pim & 0.51 & 0.17 & 12.68 & 0.03 &1.63 \cr
\Bs \to \KS \Kp \Km  & 0.09  & 0.02& 0.03 & 1.67 & 0.24 \cr
\Bs \to \KS K^{\pm} \pi^{\mp} & 4.83 & 2.10& 1.63& 0.24  &42.81 \cr
}
\times 10^{-5} 
$
}
\end{table}

\subsection{Correlation matrices due to the efficiency binning scheme}

Among the considered sources of systematic correlation, only the contribution associated with the efficiency-binning scheme is found to be significant. 
All other sources can be considered to be negligible. 
Since similar resonances appear in the different decays, a change in the Dalitz-plot efficiency binning might induce coherent shifts in the efficiency ratios of different signal modes, leading to correlated variations in the corresponding branching-fraction observables.
In addition, the variation in the normalisation channel is common to all the ratios.

The correlations among the different signal modes due to the binning of the efficiency variation over the Dalitz plot are evaluated from the changes induced when the binning
scheme is modified.
For each signal mode an efficiency ratio with respect to the reference channel, \BdtoKsPiPi, is formed in every alternative binning, data-taking period, and \KS reconstruction category, and the deviation of this ratio with respect to the baseline $10\times 10$ binning is considered.
For each pair of modes, all these deviations (over all binning schemes and subsamples) are treated as a sample from which the Pearson correlation coefficient is computed; these values are used as the correlation coefficients for the binning systematic and are reported in Tables~\ref{tab:syst-correlations-tight} and \ref{tab:syst-correlations-looser} for the tight and loose selections, respectively.
The statistical uncertainties on the coefficients are obtained from the standard Fisher $z$-transform and quoted as $\rho_{ij} \pm \sigma_{\rho_{ij}}$.
For modes that differ only by the kaon-pion charge assignment, the two-charge configurations are first averaged, so that the matrices are expressed in terms of combined $\KS K\pi$ observables.

Finally, to obtain the covariance matrices from the correlation matrices in Tables~\ref{tab:syst-correlations-tight} and \ref{tab:syst-correlations-looser}, the binning-scheme systematic uncertainties propagated to the combined branching fraction ratios are used. The resulting covariance matrices for the tighter and looser optimizations are reported in Tables~\ref{tab:syst-covariance-tighter} and \ref{tab:syst-covariance-looser}, respectively.

To obtain the systematic covariance between two branching fractions computed in the looser selection (e.g. $\Bd \to \KS \Kp \Km$ and $\Bs \to \KS K^{\pm}\pi^{\mp}$), the matrix in Table~\ref{tab:syst-covariance-looser} should be used. If both branching fractions are computed in the tighter selection (e.g. $\Bd \to \KS K^{\pm}\pi^{\mp}$ and $\Bs \to \KS \Kp \Km$), the matrix in Table~\ref{tab:syst-covariance-tighter} should be used. To obtain the covariance between one branching fraction computed in the looser selection and another computed in the tighter selection (e.g. $\Bd \to \KS \Kp \Km$ and $\Bs \to \KS \Kp \Km$), the matrix in Table~\ref{tab:syst-covariance-tighter} is used as an approximation for simplicity.
\begin{table}[tbp]
\centering
\caption{Correlation matrix due to the efficiency binning scheme for the tighter selection between the ratios of efficiencies $\eps^{\rm sel}_{\BdstoKshhp}/\eps^{\rm sel}_{\BdtoKsPiPi}$, where the numerator decay mode is given by the row and column headings.}
\label{tab:syst-correlations-tight}
\resizebox{\textwidth}{!}{
$
\bordermatrix{
    & \Bd \to \KS \Kp \Km
    & \Bd \to \KS K^{\pm} \pi^{\mp}
    & \Bs \to \KS \pip \pim
    & \Bs \to \KS \Kp \Km
    & \Bs \to \KS K^{\pm} \pi^{\mp} \cr
\Bd \to \KS \Kp \Km
    & 1.00
    & 0.40 \pm 0.09
    & 0.09 \pm 0.11
    & 0.26 \pm 0.10
    & 0.33 \pm 0.10 \cr
\Bd \to \KS K^{\pm} \pi^{\mp}
    & 0.40 \pm 0.09
    & 1.00
    & -0.14 \pm 0.11
    & -0.04 \pm 0.11
    & 0.53 \pm 0.08 \cr
\Bs \to \KS \pip \pim
    & 0.09 \pm 0.11
    & -0.14 \pm 0.11
    & 1.00
    & -0.19 \pm 0.11
    & -0.05 \pm 0.11 \cr
\Bs \to \KS \Kp \Km
    & 0.26 \pm 0.10
    & -0.04 \pm 0.11
    & -0.19 \pm 0.11
    & 1.00
    & 0.34 \pm 0.10 \cr
\Bs \to \KS K^{\pm} \pi^{\mp}
    & 0.33 \pm 0.10
    & 0.53 \pm 0.08
    & -0.05 \pm 0.11
    & 0.34 \pm 0.10
    & 1.00 \cr
}
$
}
\end{table}

\begin{table}[tbp]
\centering
\caption{Correlation matrix due to the efficiency binning scheme for the looser selection between the ratios of efficiencies $\eps^{\rm sel}_{\BdstoKshhp}/\eps^{\rm sel}_{\BdtoKsPiPi}$, where the numerator decay mode is given by the row and column headings.}
\label{tab:syst-correlations-looser}
\resizebox{\textwidth}{!}{
$
\bordermatrix{
    & \Bd \to \KS \Kp \Km
    & \Bd \to \KS K^{\pm} \pi^{\mp}
    & \Bs \to \KS \pip \pim
    & \Bs \to \KS \Kp \Km
    & \Bs \to \KS K^{\pm} \pi^{\mp} \cr
\Bd \to \KS \Kp \Km
    & 1.00
    & 0.49 \pm 0.08
    & 0.05 \pm 0.11
    & 0.22 \pm 0.11
    & 0.50 \pm 0.08 \cr
\Bd \to \KS K^{\pm} \pi^{\mp}
    & 0.49 \pm 0.08
    & 1.00
    & -0.16 \pm 0.11
    & -0.06 \pm 0.11
    & 0.65 \pm 0.06 \cr
\Bs \to \KS \pip \pim
    & 0.05 \pm 0.11
    & -0.16 \pm 0.11
    & 1.00
    & -0.17 \pm 0.11
    & -0.09 \pm 0.11 \cr
\Bs \to \KS \Kp \Km
    & 0.22 \pm 0.11
    & -0.06 \pm 0.11
    & -0.17 \pm 0.11
    & 1.00
    & 0.37 \pm 0.10 \cr
\Bs \to \KS K^{\pm} \pi^{\mp}
    & 0.50 \pm 0.08
    & 0.65 \pm 0.06
    & -0.09 \pm 0.11
    & 0.37 \pm 0.10
    & 1.00 \cr
}
$
}
\end{table}

 \begin{table}[tbp]
 \centering
 \caption{Covariance matrix due to the efficiency binning scheme for the tighter selection between the ratios of efficiencies $\eps^{\rm sel}_{\BdstoKshhp}/\eps^{\rm sel}_{\BdtoKsPiPi}$, where the numerator decay mode is given by the row and column headings.}
 \label{tab:syst-covariance-tighter}
 \resizebox{\textwidth}{!}{$
 \bordermatrix{
     & \Bd \to \KS \Kp \Km & \Bd \to \KS K^{\pm} \pi^{\mp}&  \Bs \to \KS \pip \pim& \Bs \to \KS \Kp \Km & \Bs \to \KS K^{\pm} \pi^{\mp}   \cr
 \Bd \to \KS \Kp \Km & 16.83 & 1.16 & 0.50 & 0.35 & 10.35 \cr
 \Bd \to \KS K^{\pm} \pi^{\mp} & 1.16 & 0.50 & -0.14 & -0.01 & 2.88 \cr
 \Bs \to \KS \pip \pim & 0.50 & -0.14 & 1.87 & -0.09 & -0.49 \cr
 \Bs \to \KS \Kp \Km & 0.35 & -0.01 & -0.09 & 0.11 & 0.88 \cr
 \Bs \to \KS K^{\pm} \pi^{\mp} & 10.35 & 2.88 & -0.49 & 0.88 & 58.86 \cr
 }
 \;\times 10^{-5}
 $}
 \end{table}

 \begin{table}[tbp]
 \centering
 \caption{Covariance matrix due to the efficiency binning scheme for the looser selection between the ratios of efficiencies $\eps^{\rm sel}_{\BdstoKshhp}/\eps^{\rm sel}_{\BdtoKsPiPi}$, where the numerator decay mode is given by the row and column headings.}
 \label{tab:syst-covariance-looser}
 \resizebox{\textwidth}{!}{
 $
 \bordermatrix{
     & \Bd \to \KS \Kp \Km & \Bd \to \KS K^{\pm} \pi^{\mp}&  \Bs \to \KS \pip \pim& \Bs \to \KS \Kp \Km & \Bs \to \KS K^{\pm} \pi^{\mp}   \cr
 \Bd \to \KS \Kp \Km & 15.14 & 1.09 & 0.25 & 0.37 &  9.70 \cr
 \Bd \to \KS K^{\pm} \pi^{\mp} & 1.09 & 0.33 & -0.11 & -0.01 & 1.84 \cr
 \Bs \to \KS \pip \pim & 0.25 & -0.11 & 1.57 & -0.09 & -0.55 \cr
 \Bs \to \KS \Kp \Km  & 0.37  & -0.01 & -0.09 & 0.18 & 0.77 \cr
 \Bs \to \KS K^{\pm} \pi^{\mp} & 9.70 & 1.84 & -0.55 & 0.77 & 24.72 \cr
 }
 \;\times 10^{-5}
 $}
 \end{table}

\clearpage

\addcontentsline{toc}{section}{References}
\bibliographystyle{LHCb}
\bibliography{main,standard,LHCb-PAPER,LHCb-CONF,LHCb-DP,LHCb-TDR}

\newpage
\centerline
{\large\bf LHCb collaboration}
\begin
{flushleft}
\small
R.~Aaij$^{38}$\lhcborcid{0000-0003-0533-1952},
A.S.W.~Abdelmotteleb$^{57}$\lhcborcid{0000-0001-7905-0542},
C.~Abellan~Beteta$^{51}$\lhcborcid{0009-0009-0869-6798},
F.~Abudin\'en$^{57}$\lhcborcid{0000-0002-6737-3528},
T.~Ackernley$^{61}$\lhcborcid{0000-0002-5951-3498},
A.A.~Adefisoye$^{69}$\lhcborcid{0000-0003-2448-1550},
B.~Adeva$^{47}$\lhcborcid{0000-0001-9756-3712},
M.~Adinolfi$^{55}$\lhcborcid{0000-0002-1326-1264},
P.~Adlarson$^{82}$\lhcborcid{0000-0001-6280-3851},
H.~Afsharnia$^{11}$
% ...

{\footnotesize \it

% ...
$^{11}$Universit{\'e} Clermont Auvergne, CNRS/IN2P3, LPC, Clermont-Ferrand, France\\
% ...
$^{38}$Nikhef National Institute for Subatomic Physics, Amsterdam, Netherlands\\
% ...
$^{47}$Instituto Galego de F{\'\i}sica de Altas Enerx{\'\i}as (IGFAE), Universidade de Santiago de Compostela, Santiago de Compostela, Spain\\
% ...
$^{51}$Physik-Institut, Universit{\"a}t Z{\"u}rich, Z{\"u}rich, Switzerland\\
% ...
$^{55}$H.H. Wills Physics Laboratory, University of Bristol, Bristol, United Kingdom\\
% ...
$^{57}$Department of Physics, University of Warwick, Coventry, United Kingdom\\
% ...
$^{60}$School of Physics and Astronomy, University of Glasgow, Glasgow, United Kingdom\\
$^{61}$Oliver Lodge Laboratory, University of Liverpool, Liverpool, United Kingdom\\
% ...
$^{69}$Syracuse University, Syracuse, NY, United States\\
% ...
$^{82}$Department of Physics and Astronomy, Uppsala University, Uppsala, Sweden, associated to $^{60}$\\
% ...
}
\end{flushleft}

\end{document}